\documentclass[journal]{IEEEtranTCOM}%

\usepackage[pdftex]{graphicx}
\usepackage{epstopdf}
\usepackage{amsmath}
\usepackage{cite}
\usepackage{amssymb}
\usepackage{booktabs}
\usepackage{array, makecell}
\usepackage{multirow}

\def\BibTeX{{\rm B\kern-.05em{\sc i\kern-.025em b}\kern-.08em
T\kern-.1667em\lower.7ex\hbox{E}\kern-.125emX}}

\DeclareMathAlphabet\mathbfcal{OMS}{cmsy}{b}{n}
\DeclareMathAlphabet{\mathbfscr}{OMS}{mdugm}{b}{n}
\let\originalleft\left
\let\originalright\right
\renewcommand{\left}{\mathopen{}\mathclose\bgroup\originalleft}
\renewcommand{\right}{\aftergroup\egroup\originalright}

\title{Scheduling Versus Contention for Massive Random Access in Massive MIMO Systems}

\begin{document}

\author{Justin Kang,~\IEEEmembership{Graduate Student Member, IEEE}, and
Wei Yu,~\IEEEmembership{Fellow, IEEE}%
\thanks{
	Manuscript submitted to IEEE Transactions on Communications on December 31, 2021; revised May 12, 2022; accepted July 4, 2022.
This work was supported by Natural Sciences and Engineering Research Council (NSERC) of Canada via
a discovery grant.

Justin Kang was with The Edward S. Rogers Sr. Department of Electrical and Computer Engineering, University of Toronto, Toronto, ON M5S
3G4, Canada. He is now with the Department of Electrical Engineering
and Computer Science, University of California, Berkeley, CA 94720 USA.
(e-mail: justin{\textunderscore}kang@eecs.berkeley.edu).

Wei Yu is with The Edward S. Rogers Sr. Department of Electrical 
and Computer Engineering, University of Toronto, Toronto, ON M5S 3G4, Canada.
(e-mail: weiyu@ece.utoronto.ca).}%
}

\maketitle

\begin{abstract}
Massive machine-type communications protocols have typically been designed
under the assumption that coordination between users requires significant
communication overhead and is thus impractical.  Recent progress in efficient
activity detection and collision-free scheduling, however, indicates that 
the cost of coordination can be much less than the naive scheme for
scheduling.  This work considers a scenario in which a massive number of
devices with sporadic traffic seek to access a massive multiple-input
multiple-output (MIMO) base-station (BS) and explores an approach in which
device activity detection is followed by a single common feedback broadcast
message, which is used both to schedule the active users to different
transmission slots and to assign orthogonal pilots to the users for channel estimation. 
The proposed coordinated communication scheme is compared to two prevalent
contention-based schemes: coded pilot access, which is based on the
principle of coded slotted ALOHA, and an approximate message passing scheme
for joint user activity detection and channel estimation.  
Numerical results indicate that scheduled massive access provides
significant gains in the number of successful transmissions per slot and in sum rate,
due to the reduced interference, at only a small cost of feedback.
\end{abstract}
\IEEEpeerreviewmaketitle


\section{Introduction} \label{sec:introduction}
Future wireless networks need to support massive connectivity in the form of the Internet of Things (IoT) and massive machine-type communications (mMTC). In a massive connectivity scenario, a single cellular base station (BS) must support a large number of $N$ devices (in the order of $10^4 \sim 10^6$). A salient characteristic of IoT and mMTC traffic is that devices typically seek to access the network only \textit{sporadically} and only to transmit small payloads, so that at any given time only a small random subset of $K \ll N$ users are active \cite{XChenCapacityManyAccess, ChenMassiveAccess}. In this setting it is highly inefficient to pre-assign each user an orthogonal communication resource, since the user only sporadically makes use of it. Designing solutions for massive connectivity that allow users to efficiently gain {random access} to the network thus becomes an important problem. 

Existing wireless protocols \cite{LTE2013} already implement random access, but on a much smaller scale than is envisioned for massive connectivity in IoT and mMTC \cite{Wu2020}. Most of these existing protocols are based on \textit{contention}. For example, in the Long Term Evolution (LTE) standard, users request access for available resources at random; the BS then transmits a downlink feedback acknowledgement message, which is followed by an uplink response from the users and finally another downlink feedback from the BS for resolving any colliding resource requests and to authenticate the users. 

With the goal of designing scalable random access solutions to meet the requirements of future systems, many random access schemes that differ from the above traditional approach have been proposed \cite{Sorensen2018, Fengler2021}. Among these proposed schemes, the \textit{grant-free} paradigm \cite{Liu2018SparseThings} is popular. Grant-free protocols rely entirely on uplink communications and focus on dealing with the inevitable interference resulting from user contention.
 This is typically justified by the fact that the alternative, i.e., using downlink feedback to enable scheduling in order to eliminate interference would come at too great a cost \cite{Sorensen2018}.

Recent discoveries in two separate areas, however, have shown that the cost of downlink feedback for scheduling to avoid interference may be less than previously thought.
First, for the massive multiple-input multiple-output (MIMO) system, compressed sensing algorithms such as {\it approximate message passing} (AMP) can be used for accurate detection of the active users \cite{Liu2018_1} and to simultaneously estimate their channels \cite{Liu2018}. 
Moreover, if only the large-scale fading needs to be estimated, 
then it is possible to detect $K = O \left(L^2 \right)$ active devices with only $L$ pilot symbols using a technique known as the \textit{covariance approach} \cite{Haghighatshoar2018}. 

The second discovery is that after activity detection, coordination among the active devices can be enabled via a common feedback message from the BS to the active users and that the amount of feedback required to ensure  collision-free scheduling scales only linearly in $K$ with a coefficient as small as $1.44$ bits per active user, and nearly independent of $N$, in theory. The feedback cost is even less if multiple users can be scheduled in the same time or frequency slot \cite{kang2020minimum}.


Together, these two sets of results suggest the following three-step procedure for massive
connectivity with massive MIMO. In the first stage, active users transmit
uniquely identifying non-orthogonal pilots; the BS performs sparse activity
detection based on compressed sensing. In the second stage, the BS transmits 
a common feedback message to the active users. In the final phase, the active
users transmit additional pilots for channel estimation as well as the payload,
while making use of the feedback message both for assigning pilots and for scheduling data
transmission into orthogonal slots, in order to avoid interference. This scheme differs
from grant-free schemes by focusing on the prevention of interference, rather
than mitigating its effects. 

The main goal of this paper is to show the significant throughput improvement
that can be obtained for the scheduled scheme as compared to the contention
scheme for massive random access, and that such benefit comes only at a
small cost of feedback.

\subsection{Related Work}

The classic strategy for implementing contention-based random access is {Slotted ALOHA} (SA) \cite{Roberts1975ALOHACapture}. In classic SA users randomly transmit in orthogonal slots and re-transmit in the case of collision. 
In these systems, the largest fraction of orthogonal slots that can be effectively utilized for user transmission is ${1}/{e}$, resulting in a significant waste of resources. There are many modern variations of SA that seek to remedy this by including redundancy in the user transmission (typically through the repeated transmission of the payload) and by utilizing information from collisions via successive interference cancellation (SIC). Most of these methods \cite{Casini2007, Liva2011Graph-BasedALOHA, Narayanan2012} fall into the category of {Coded Slotted ALOHA}  (CSA) \cite{Paolini2015}. These schemes can utilize a much higher fraction of the available resources. In particular, the scheme presented in \cite{Narayanan2012} exploits a connection to the erasure decoding in fountain codes. 
By making use of the soliton distribution in the design of collision resolution code, it is shown that CSA can asymptotically approach perfect utilization as the number of slots and users approach infinity. This advantage however comes at a cost, as these schemes often require a long block length and require the users to transmit the same packet multiple times, resulting in additional energy consumption. 

The CSA protocol describes random access at a packet level, ignoring the underlying physical layer. To utilize CSA methods in a practical setting, one must also account for the physical layer transmission concerns. In particular, massive MIMO \cite{Marzetta2010}, where each BS is equipped with a large number of antennas, has emerged as a key technology for future wireless systems, making random access for massive MIMO an important research direction. Critically, the design of massive MIMO systems must address the important issue of channel estimation. Toward this end, \cite{Sorensen2018} introduces a random access protocol for massive MIMO known as {Coded Pilot Access} (CPA), which uses randomly selected orthogonal pilots for channel estimation, as well as the concepts of CSA to resolve collisions. CPA is a benchmark against which the methods proposed in this paper are compared.


%

In contrast to the contention-based strategies, this paper explores
alternatives that are based on the scheduling of the active users in orthogonal
slots.  Conventionally, scheduling $K$
users out of a potential pool of $N$ users would require a feedback message of $K\log(N)$ bits.  Surprisingly, in \cite{kang2020minimum}, it is
revealed that if only $K$ active users out of $N$ total users are
listening to the feedback message, and each active user is only interested in
knowing its own scheduled slot, then the fundamental bounds on the size of the
common feedback message required to ensure collision-free scheduling can be
much smaller. Information theoretically, it is shown in \cite{kang2020minimum}
that scheduling $K$ users into $K$ slots while avoiding collision only 
requires approximately $\log(e) K$ bits of common feedback, plus an additive
term that scales as $O(\log\log(N))$ if fixed-length code is used. The fact
that the optimal collision-free feedback can be highly efficient is a main
motivation for the present work.

The use of scheduling and feedback for massive connectivity has already been 
considered in several recent works \cite{Facenda2020, Romanov2021}, 
but for a different context of \emph{unsourced random access}
\cite{Polyanskiy2017ARandom-access, amalladinne2018coded, fengler2021pilotbased},
where user identification is abstracted, and the goal is to decode 
a list of transmitted messages. 
The unsourced paradigm is most suitable when the messages themselves, rather 
than the identities of the transmitters, are important.

This present paper considers the \emph{sourced} approach, in which the BS is
made aware of the identities of the active users through an activity detection process, then uses a feedback strategy
to schedule the active users. In the activity detection process, each user is assigned a unique signature sequence. Due to the large number of devices in the user pool, the signature sequences cannot be orthogonal. But because of the sporadic nature of the device activities, compressed sensing techniques can be used to recover the identities of the active users.  In
\cite{Liu2018_1, Chen2018SparseConnectivity, Ke2020} the AMP algorithm is proposed for activity detection in multi-antenna systems. Importantly,
the performance of these activity detection methods that employ AMP can be
predicted by an analytic framework called state evolution \cite{Donoho2009}. 
The AMP algorithm works by performing joint user activity detection and
instantaneous channel state information (CSI) estimation. We note however that when the
number of antennas is large, the problem of instantaneous CSI estimation from
non-orthogonal pilots becomes more difficult, and the convergence of the AMP
algorithm becomes considerably slower. 

 In \cite{Haghighatshoar2018}, an alternative approach to activity detection is considered based on the key insight that the sample covariance matrix of the received signal is a sufficient statistic for detecting the active users. This approach, known as the \textit{covariance approach}, forgoes CSI estimation and has several advantages when the number of antennas is large. In \cite{Fengler2021, Chen2019phase} the performance of the covariance approach is studied asymptotically via a phase transition analysis and numerically for finite parameters, showing that in the massive MIMO setting, it outperforms AMP.
 
 Additionally, the idea of joint activity and data detection is considered in \cite{Senel2018}. To achieve this, each user is assigned multiple signature sequences and selects one based on the data it wishes to transmit. This is advantageous because it does not require any coordination and can be implemented as a straightforward extension of activity detection. However, since the total number of required sequences grows exponentially with the size of the data payload, it is suitable only for very small payloads.

\subsection{Main Contributions}

This paper studies the use of feedback to improve massive random access schemes in a massive MIMO system. We allow for a single common feedback message from the BS to the users to enable scheduling. The main contributions are as follows:
\begin{itemize}
\item{We propose a three-phase random access scheme that exploits activity detection and feedback to enable coordination. In the first phase, active users transmit non-orthogonal uplink pilots for activity detection. After the BS has determined the set of active users, it broadcasts a common downlink feedback message to assign each user a slot and an orthogonal pilot. Importantly, we utilize this feedback not just to schedule users, but also to assign orthogonal pilots, which resolves the critical issue of channel estimation in massive MIMO design. In the final phase, the users transmit their orthogonal pilots and data in the scheduled slot.}
\item{For the case where the communications occur over multiple coherence blocks, we compare the proposed scheduled approach to random access with contention-based CPA \cite{Sorensen2018}.  Numerical results indicate that the significant performance gains in terms of number of successful transmissions and in system efficiency can be obtained at a cost of only a small amount of feedback.}
\item{
For the case where the communications occur over a single coherence block, we first
show that the feedback rate required for scheduling users into different
transmission slots as proposed in \cite{Liu2018} is very small, then show that 
if a moderately higher feedback rate is used to allocate orthogonal 
pilots for channel estimation, then additional gains in system sum rate can be 
obtained as compared to the AMP-based joint user activity detection and channel 
estimation scheme described in \cite{Liu2018}. 
}
\end{itemize}
Together, these results quantify the benefit and the cost of scheduling
orthogonal resources for both channel estimation and data transmission in
massive random access for massive MIMO systems.

\subsection{Organization}
The rest of the paper is organized as follows. Section \ref{sec:formulation}
presents the system model and the problem formulation. In Section
\ref{sec:activity}, we review two compressed sensing algorithms for device
activity detection for massive random access. In Section \ref{sec:feedback},
fundamental bounds for collision-free feedback scheduling are presented. We
then propose the coordinated random access scheme that uses common feedback
from the BS to the users for scheduling and for pilot assignment for channel
estimation, and numerically compare its performance with the uncoordinated
contention-based alternatives for the fast-fading scenario in Section
\ref{sec:fast_fading} and for the slow-fading scenario in Section
\ref{sec:slow_fading}. The paper concludes with Section \ref{sec:conclusion}.


\subsection{Notation}
Throughout the paper standard upper and lower-case symbols denote scalars. Lower-case and upper-case boldface symbols denote vectors and matrices respectively. Calligraphy letters denote sets. Superscripts $(\cdot)^{T}$ and $(\cdot)^H$ denote transpose and conjugate transpose respectively. Further, $\mathbf{I}$ represents the identity matrix with appropriate dimensions, and $\mathcal{CN}(\boldsymbol{\mu}, \mathbf{\Sigma})$ denotes a complex Gaussian distribution with mean $\boldsymbol{\mu}$ and covariance $\mathbf{\Sigma}$. The set $[N]$ denotes the set $\{1, \dotsc, N\}$ and $|\cdot|$ denotes the number of elements of a set. All logarithms are base $2$ unless otherwise stated.

\section{Problem Formulation}\label{sec:formulation}

Consider the uplink of an mMTC system consisting of a single BS with $M$
antennas, and $N$ potential users with a single antenna each. Communications
occur over a \textit{frame}, corresponding to the time scale in which
the users' activities are fixed, and can also be thought of as the latency constraints
within which the active users must be served. We assume that among a large
number of $N$ potential users, a random subset of $K$ users are active and 
seek to transmit a small payload to the BS. Let $\mathcal{A} \subset [N]$ with
$|\mathcal{A}| = K$ denote the set of indices of the active users.   

\begin{figure}[t]
\centering
\ifdefined\ONECOLUMN
\includegraphics[width=0.6\columnwidth]{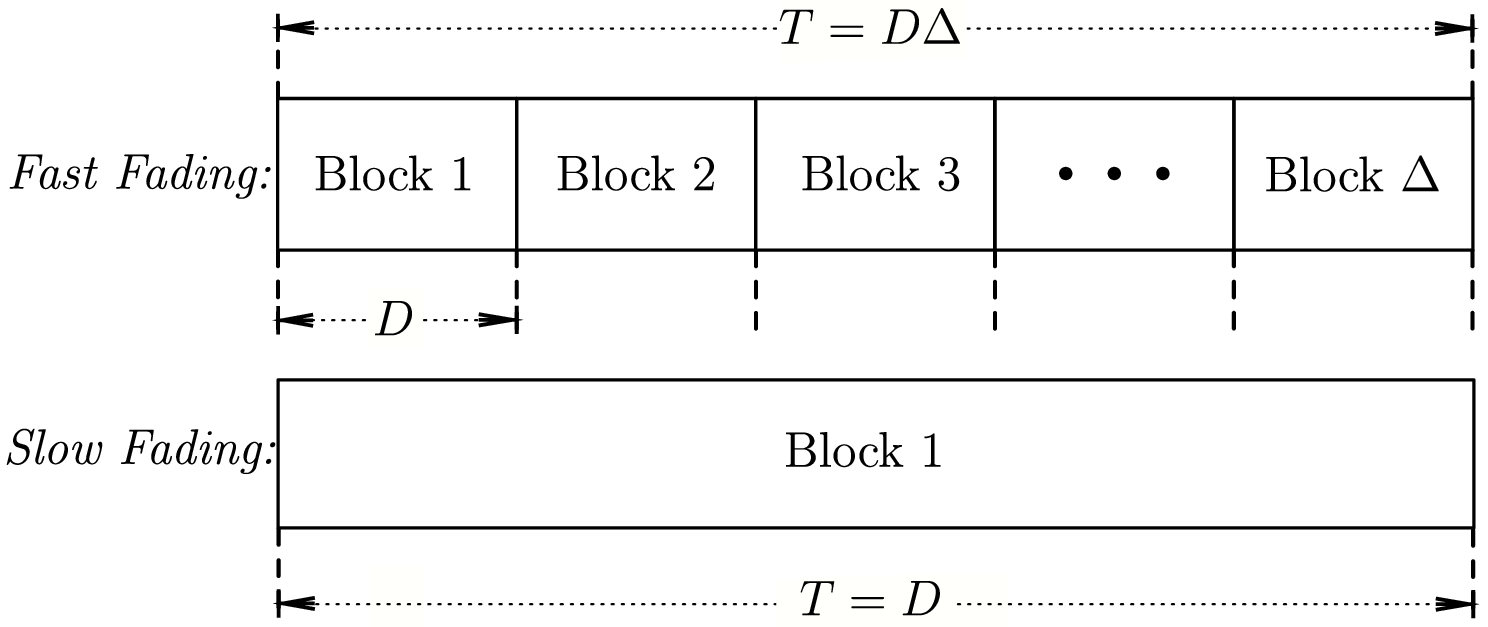}
\else
\includegraphics[width=0.9\columnwidth]{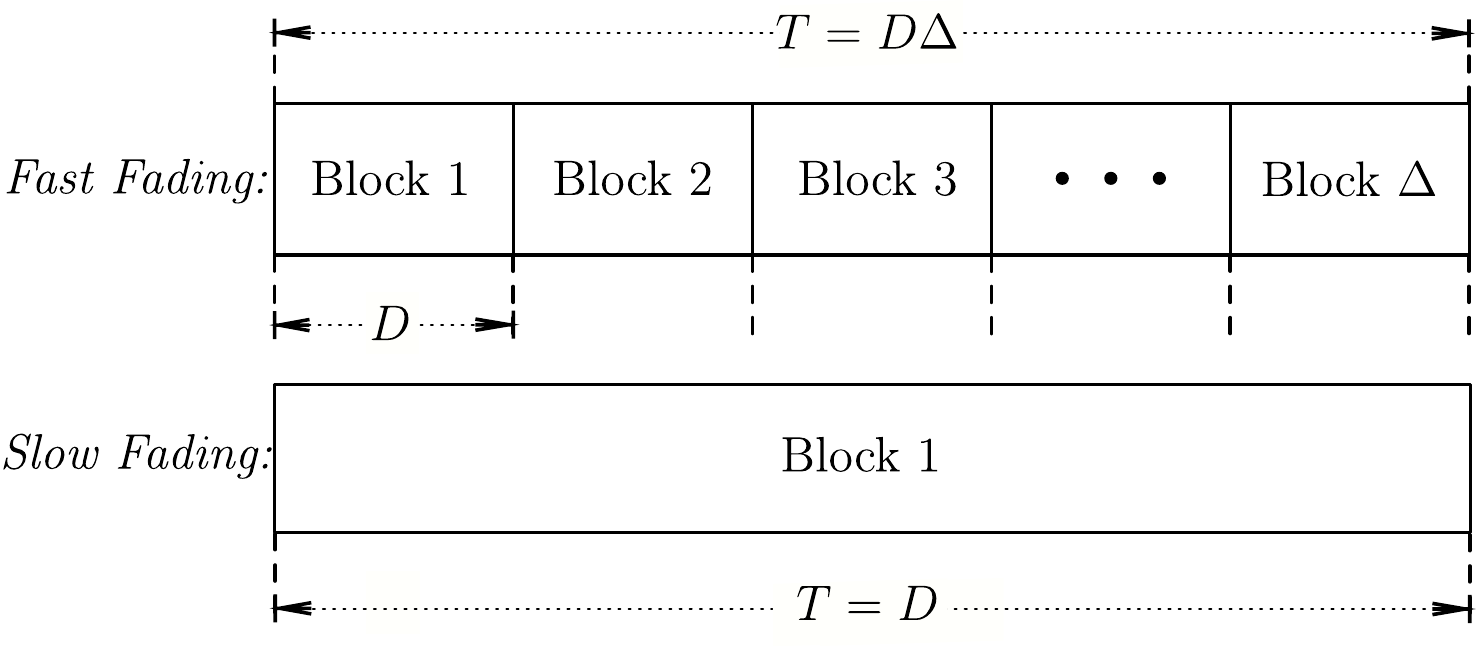}
\fi
\caption{Relationship between frame length $T$ and coherence length $D$ for the fast-fading and slow-fading channel models.}
\label{fig:problem_setup}
\end{figure}

The uplink channels are modelled as an independently and identically distributed
(i.i.d.) block-fading wireless channel, where the users' small-scale fading
coefficients remain stable for a fixed coherence block. We consider two
different scenarios: 
\begin{enumerate}
\item[(i)] In the \emph{fast-fading} 
scenario, the coherence block is shorter than the frame length, and
each frame can be thought of as consisting of $\Delta$ consecutive coherence 
blocks, each of length $D$ channel uses, resulting in a total of $T = D \Delta$ 
channel uses per frame. 
\item[(ii)] In the \emph{slow-fading} scenario, the coherence
length is longer than the frame length, so that without loss of generality, we 
can assume $\Delta=1$ and $T=D$.
\end{enumerate}
Fig.~\ref{fig:problem_setup} illustrates the relations between the frame length and the coherence length for the two cases. Note that the user activity detection needs to take place within each frame length $T$, while channel estimation needs to take place within each coherent length $D$. Note that the block length corresponding to coding and modulation would typically be much smaller than $D$, i.e., each coherence block would consist of many transmission symbols. 

\begin{table*}[h]
\centering
\caption{ Summary of Problem Formulations and Assumptions}
\resizebox{!}{!}{%
\begin{tabular}{l l l}
\toprule
 Fading Model &  Benchmark Protocol & Proposed Scheduled Approach \\ 
\midrule
Fast Fading &  Coded Pilot Access \cite{Sorensen2018} & \makecell[l] {   Activity Detection, Feedback \\    and Scheduled Transmission} \\
\midrule
Slow Fading & \makecell[l]{   Joint Activity and Channel \\  Estimation with Non-Orthogonal \\    Pilots Using AMP \cite{Liu2018}} & \makecell[l]{   Activity Detection with Covariance \\    Method and Scheduled Orthogonal \\    Pilots for Channel Estimation } \\
\bottomrule
& & \\
\end{tabular}
}
\label{tab:problem_sum}
\end{table*}

The fading channel between the $i$th user and the BS in the $d$-th coherence block 
is denoted as $g_i \mathbf{h}_{d,i} = g_i \left[ h_{d,i}(1)\;
h_{d,i}(2)\;\dotsc\;h_{d,i}(M) \right]^T$ where $\mathbf{h}_{d,i} \in
\mathbb{C}^{M \times 1}$, $\mathbf{h}_{d,i} \sim \mathcal{CN}(0,\mathbf{I})$ is
the Rayleigh fading component and $g_i \in \mathbb{R}_+$ is the large-scale
fading. In each coherence block, some (or all) of the active users would choose or be scheduled to transmit to the BS. Let $\mathcal{A}_d \subseteq \mathcal{A}$ denote the set of transmitting users in the $d$-th coherence block. We can write the received signal at the
BS in the coherence block $\mathbf{Y}_d \in \mathbb{C}^{M \times D}$
as: \begin{equation}
\mathbf{Y}_d = \sum _{j \in \mathcal{A}_d} g_j \mathbf{h}_{d,j} \mathbf{x}_{d,j}  + \mathbf{Z}_{d}, \;\; d = 1,\dotsc,\Delta,
\end{equation}
 where $\mathbf{x}_{d,j}$ is the signal transmitted by user $j \in \mathcal{A}_d$, and $\mathbf{Z}_d \in \mathbb{C}^{M \times D}$ is the additive white Gaussian noise (AWGN) with i.i.d.\ elements distributed according to $\mathcal{CN}(0, \sigma_n^2)$.
 At the end of the frame, the BS would use all $\mathbf{Y}_d$, $d = 1, \dotsc, \Delta$ to determine the set of active users and their associated payloads. 
 Note that this formulation is distinct from unsourced random access \cite{Polyanskiy2017ARandom-access}, for which only a list of payloads is required to be decoded. 

For protocols that involve feedback, we assume that it takes the form of a single common broadcast message from the BS to the users. The feedback message occurs at some point during the frame. When $\Delta > 1$, we assume that feedback occurs between the coherence blocks, and in the slow-fading model with $\Delta=1$, we assume that feedback occurs within the coherence block. In this work, we avoid modelling the physical feedback channel, and instead quantify the cost of feedback through a characterization of the minimum amount of information (i.e., number of bits) required to be broadcast in order to achieve the scheduling objective.

We consider the two distinct fading models in order to compare the proposed scheduled scheme with existing random access protocols. A summary of the models along with the existing and proposed new protocols is presented in Table~\ref{tab:problem_sum}.

The assumptions made above falls in line with most works on massive MIMO. We remark that the massive MIMO system has also been studied under more realistic propagation conditions involving correlation between channels and partial line-of-sight propagation \cite{Zhang2013,Gao2015}. For consistency and to capture the fundamental aspect of the problem, this paper considers the i.i.d.\ fading model only. A study of the benefits of feedback and scheduling in models with correlation is left to future work.

\section{Sparse Activity Detection}\label{sec:activity}

Activity detection is the process by which the BS determines the identities of the $K$ active users among the $N$ potential users in each frame.  There are two well-known approaches, one using the AMP algorithm, and the other, based on a covariance estimation formulation. Both approaches have been shown to be theoretically and practically viable under a range of system parameters. Throughout this work, we assume to operate in regimes where activity detection is feasible.

We take a \emph{sourced} random access approach in which each of the $N$ potential users are assigned uniquely identifying non-orthogonal pilot sequences  $\mathbf{s}_1, \mathbf{s}_2, \dotsc, \mathbf{s}_n \in \mathbb{C}^{L}$. In the pilot phase, the active users in $\mathcal{A}$ transmit their pilots to the BS. In this case, the received signal can be expressed as:
\begin{eqnarray}
\mathbf{Y}^T &=& \sum_{i \in \mathcal{A}} \mathbf{s}_i g_i \mathbf{h}^T_{i}   + \mathbf{Z} \\
&=&  \mathbf{S} \boldsymbol{\Gamma}^{\frac{1}{2}} \mathbf{H} + \mathbf{Z}, \label{eq:AD_mtrx_form}
\end{eqnarray}
where $\mathbf{Y} \in \mathbb{C}^{M\times L}$ is the received signal, $\mathbf{S} \triangleq [\mathbf{s}_1, \dotsc, \mathbf{s}_N] \in \mathbb{C}^{L \times N}$ is the signature sequence matrix, $\boldsymbol{\Gamma} \triangleq \text{diag}\left\{\gamma_1, \dotsc, \gamma_N\right\} \in \mathbb{R}_{+}^{N \times N}$ 
where $\gamma_i = (a_i g_i)^2$ and $a_i = 1$ if $i \in \mathcal{A}$ and
otherwise $0$, and $\mathbf{H} \in \mathbb{C}^{N\times M}$ is the combined channel for all users, and $\mathbf{Z} \in \mathbb{C}^{M\times L}$ is the AWGN noise matrix.
We assume that $\mathbf{S}$ is known at the BS. 

\subsection{AMP Approach}

One way to formulate the problem of activity detection is to note that the effective  CSI matrix  $\mathbf{X} \triangleq \boldsymbol{\Gamma}^{\frac{1}{2}}\mathbf{H}$ is row-sparse, and the non-zero rows correspond to the active users. Rewriting \eqref{eq:AD_mtrx_form} as 
\begin{equation}\label{eq:AMP_mmv}
\mathbf{Y}^T =\mathbf{S} \mathbf{X}  + \mathbf{Z},
\end{equation}
the problem of estimating the user activities can now be formulated as that of estimating the sparsity pattern of $\mathbf{X}$ from the observation $\mathbf{Y}$, which can be seen as a multiple measurement vector (MMV) compressed sensing problem \cite{Ziniel2013}. 
One way to solve this problem is to use the AMP algorithm, which yields not only 
the sparsity pattern but also an estimate of the matrix $\mathbf{X}$. 
Thus, the AMP approach in fact amounts to joint sparsity activity detection and channel estimation \cite{Liu2018_1,Liu2018}.
This is useful in the slow-fading scenario, because the
instantaneous CSI remains constant within the frame so the channel 
estimated from the pilot stage can then be used to design the receiver
for data reception.
In contrast, for fast-fading case, where the data transmission occurs
in a separate coherence block from activity detection, the estimated value
for $\mathbf{X}$ would not be useful since the instantaneous CSI would have
changed and would need to be re-estimated when data transmissions occur.

\subsection{Covariance Approach}

When the instantaneous CSI is not needed and only the device activities are of 
interest, an alternative approach is to consider $\mathbf{H}$ as random, and to 
treat $ \gamma_i = a_i g_i$ as deterministic unknown parameters to be estimated. 
This approach is proposed in \cite{Haghighatshoar2018, Fengler2021}, in which 
 the maximum-likelihood estimation (MLE) of $\boldsymbol{\gamma}$ from $\mathbf{Y}$ 
is formulated as the following non-linear optimization problem:
\begin{equation} \label{eq:MLE}
\begin{aligned}
\min_{\boldsymbol{\gamma}} \quad & \log \left| \boldsymbol{\Sigma} \right| +\textrm{tr}\left( \boldsymbol{\Sigma}^{-1}\hat{\boldsymbol{\Sigma}}\right) \\
\textrm{s.t.} \quad &\boldsymbol{\gamma}\geq0, \\
\end{aligned}
\end{equation}
where $\boldsymbol{\Sigma} \triangleq \mathbf{S} \boldsymbol{\Gamma}
\mathbf{S}^H + \sigma_n^2 \mathbf{I}$ and $\hat{\boldsymbol{\Sigma}} =
\frac{1}{M} \mathbf{Y}^H \mathbf{Y}$ are in $\mathbb{C}^{L \times L}$. 
This approach is most effective in 
the massive MIMO regime, where the channel hardening effect takes place
and $\hat{\boldsymbol{\Sigma}} \rightarrow {\boldsymbol{\Sigma}}$ as $M \rightarrow \infty$. 
Although the problem (\ref{eq:MLE}) is non-convex, there are relatively simple
and highly effective algorithms for numerically finding (local optimal) solutions 
\cite{Haghighatshoar2018}.
Because this problem formulation works in the covariance domain for $\mathbf{Y}$,
it is termed \emph{covariance approach} in the literature.

\subsection{Performance}

Both the AMP and the covariance approach have a strong theoretical foundation and analysis.  For the AMP algorithm, the state evolution provides theoretical guarantees on its asymptotic performance as $L$, $N$ and $K$ go to infinity (at fixed $M$) \cite{Liu2018}. For the covariance approach, a phase transition analysis has been developed in the regime of large $M$ \cite{Fengler2021, Chen2019phase}.

While the AMP algorithm is suited for moderate $M$ and has the benefit of being
able to provide an estimate of the instantaneous CSI, the covariance approach has
a distinct advantage at large $M$, because it takes advantage of the 
channel hardening effect in the massive MIMO regime. The problem formulation
\eqref{eq:AMP_mmv} aims to detect $K \times M$ non-zero entries in $\mathbf{X}$ from
$L \times M$ observations in $\mathbf{Y}$, so the AMP algorithm is expected to
be able to estimate $\mathbf{X}$ well only in the regime where $L$ is
comparable or larger than $K$. In contrast, the problem formulation
\eqref{eq:MLE} aims to detect $K$ non-zero entries in $\boldsymbol{\Gamma}$
from $L \times L$ observations in $\boldsymbol{\Sigma}$, so it can work in the
regime where $K = O(L^2)$.  This phase transition phenomenon is rigorously
established in \cite{Fengler2021, Chen2019phase}.  Numerical simulations
comparing the activity detection performance of AMP versus covariance approach 
can be found in \cite[Figs.~9 and 10]{Chen2019phase}.

The choice of whether to use AMP versus the covariance approach in practice
depends on the system setup and the operating regime. When
$L<K$, the covariance approach would significantly outperform AMP. 
When $L>K$, the two have comparable performance, with the covariance approach
having slightly better detection error performance, but with the AMP having 
the benefit of being able to provide an estimate of the channel in addition. 
In the fast-fading scenario where the estimated channel is not useful, (because the 
data transmission would have occurred in a different coherence block), it is
preferable to use the covariance method.  In the slow-fading scenario, the AMP
has the advantage of being able to provide an estimate of the channel, but
since non-orthogonal sequences are used in the pilot phase, the channel
estimation error based on AMP alone can be large \cite{Liu2018}. 
Thus, a subsequent channel estimation stage that uses additional feedback to
assign orthogonal pilot sequences to the active users can further improve 
the performance. 

\subsection{Complexity} \label{sec:complexity}

An equally important consideration for activity detection algorithms is complexity, as practical activity detection algorithms must have a run-time comparable to the time scale of the transmission frame and remain feasible even as the number of users $N$ and the number of antennas $M$ grow large. 

In this respect, we first note that the AMP is based on the observation of an $L\times M$ matrix, while the covariance approach is based on an $L \times L$ matrix. The AMP algorithm has a manageable complexity only when $M$ is small; its convergence speed slows down considerably as $M$ increases. In the regime of large $M$, the covariance approach has a significant advantage.

\begin{figure}[t]\centering
\centering
\ifdefined\ONECOLUMN
\includegraphics[width=0.7\columnwidth]{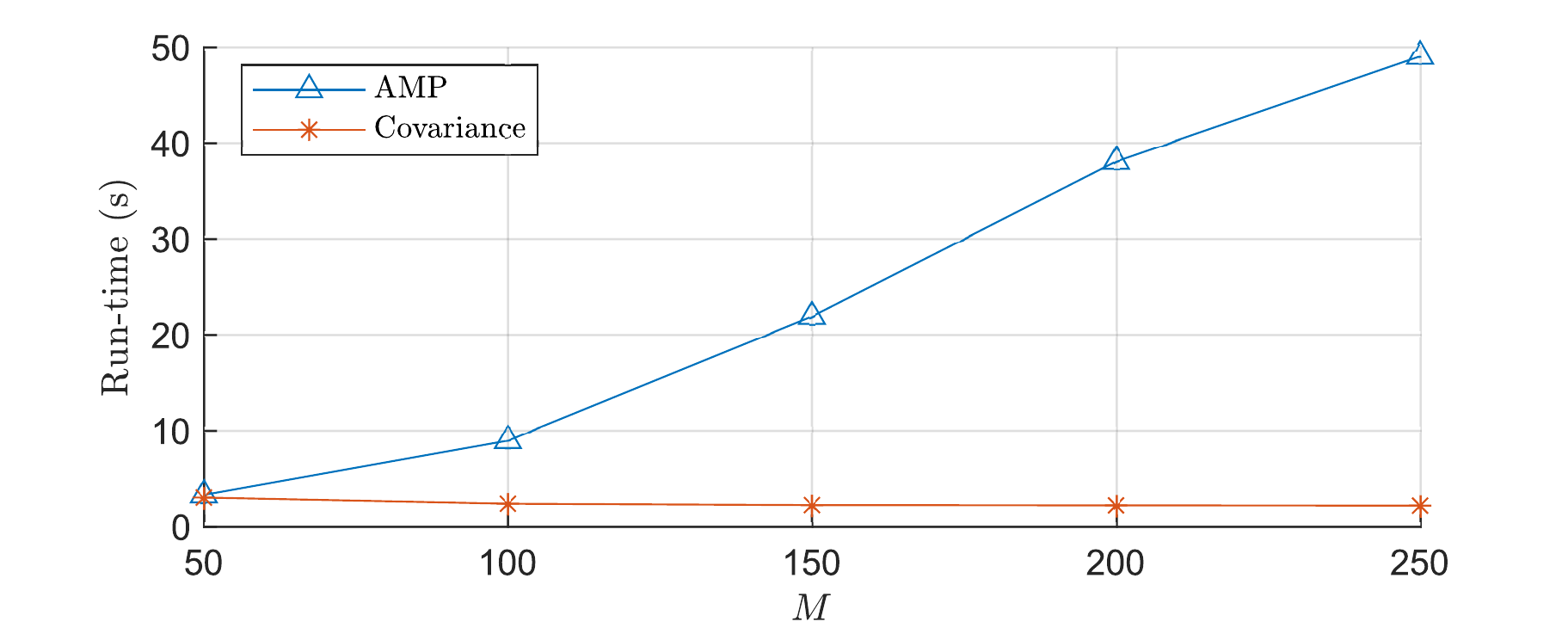}
\else
\includegraphics[width=\columnwidth]{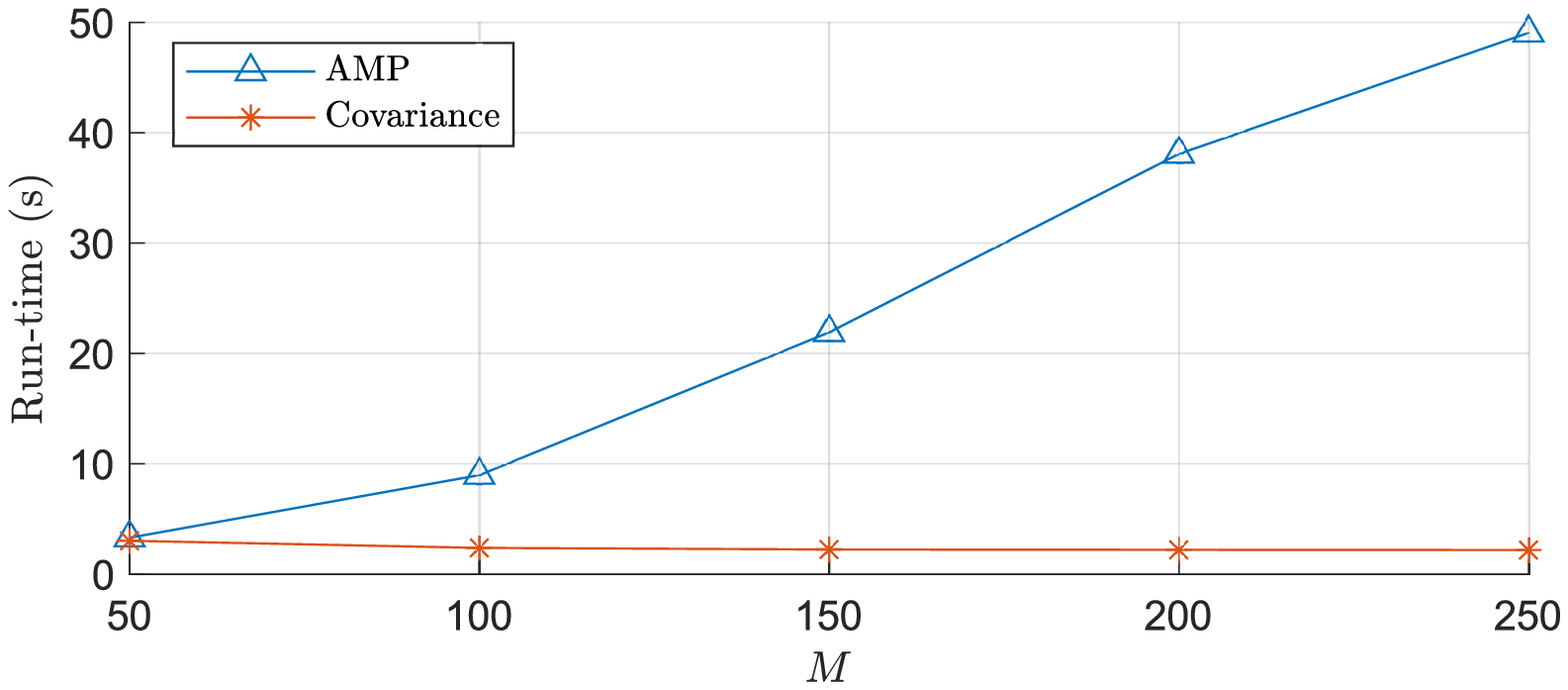}
\fi
\caption{Run-time comparison of the AMP versus the covariance approach as function of $M$, with $K=100$, $N=1000$ and $L=100$. }
\label{fig:run_time}
\end{figure}

Consider the coordinate descent algorithm for solving \eqref{eq:MLE} for the covariance approach. As noted in \cite{Chen2019phase}, the complexity of each coordinate update is $O(L^2)$. Suppose that the coordinate descent algorithm requires each of the $N$ coordinates of $\hat{\boldsymbol{\gamma}}$ to be updated $W$ times. Then, the overall complexity is $O(L^2NW)$. Note that since the covariance approach involves averaging over the antennas, the algorithm does not directly scale in complexity with $M$, making it well suited for the massive MIMO setting. 
In contrast, the MMV compressed sensing problem \eqref{eq:AMP_mmv} involves estimating the row-sparse $N \times M$ matrix $\mathbf{X}$, so the run-time of the AMP algorithm has a strong scaling with $M$. This can be observed in Fig.~\ref{fig:run_time}, which shows a run-time comparison of both algorithms using the same computing hardware with increasing $M$.

It should be noted that the expression for the complexity of the covariance approach does not necessarily imply that the complexity increases quadratically with $L$. This is because the number of iterations needed for convergence strongly depends on how close the operating point is from the phase transition boundary. 
For example, at fixed $M$, $N$ and $K$, if $L$ increases, then $W$, the number of
iterations required to converge to the optimal solution $\boldsymbol{\gamma}$, should
decrease, as the operating point is now further away from the feasibility boundary.

\section{Minimum Feedback for Collision-Free Scheduling}\label{sec:feedback}

Once the set of active users has been detected (with the result 
denoted here as $\hat{\mathcal{A}}$), the BS can then transmit a
 common feedback message to schedule the active
users into the transmission slots within the frame. A naive feedback scheme 
is to send a list of indices of the $K$ active users in the order in which
they should transmit.  This requires $K \log(N)$ bits of feedback.  It turns
out that this naive scheme is not the most efficient feedback mechanism.  
In this section, we explore the optimal collision-free feedback strategy and 
characterize the minimum feedback rate.

\subsection{Fundamental Limits of Collision-Free Feedback}

The naive feedback strategy is not the most efficient feedback mechanism 
for several reasons.
First, if the objective of the scheduling is to avoid collision, then the BS
can choose any of the $K!$ permutations of the list of
users---removing this freedom can reduce the feedback rate. This improvement
already leads to a more efficient \textit{enumerative source coding}
\cite{Facenda2020, Cover1973} method of feedback, but even this is still far
from optimal.  To approach the fundamental limit, the key observation of
\cite{kang2020minimum} is that the only information required by an active user
is which slot it is scheduled in and any information about the other users is
redundant. Thus, the full list of active users contains more information than
what each user needs for collision-free scheduling. Additionally, the naive
scheme also informs the inactive users that they are not on the list of active
users. Since inactive users are not listening to the feedback message, this
information is also redundant. These observations can be exploited when
investigating the fundamental limit of collision-free feedback.

Specifically, we define a collision-free scheduling code as the following: For any set of active users $\hat{\mathcal{A}}$ as determined by the BS (with $|\hat{\mathcal{A}}| \le K$) that need to be scheduled into $B$ slots, there must exit a codeword $\mathbf{c}$ in the feedback code of size $C$ such that:
\begin{equation}
  q_i(\mathbf{c}) \neq q_j(\mathbf{c}), \quad \forall i \neq j \in \hat{\mathcal{A}}
\end{equation}
where $q_i, i\in[N]$, are the feedback decoders for each user 
that map the user into one of the $B$ available slots.  Given a set of active users,
the output of the feedback encoder is simply the index of such a codeword.  
The rate of the feedback code is defined to be $\log(C)$ if the feedback
message must have a fixed length, or the entropy of the output of the encoder, 
if the feedback can have variable lengths.

The fundamental limit of collision-free feedback is found in 
\cite{kang2020minimum} as follows. If the number of available slots $B=K$, for the variable-length case, an achievable rate for the collision-free feedback code is $\log(e)(K+1)$ bits. Remarkably, this feedback rate is independent of $N$. The proof relies on a random set partitioning argument. In addition, converse results are also available, which indicate that for sufficiently large $N$ and $K$, this bound is tight to within $\log(e)$ bits. For the case of fixed-length feedback codes, it can be shown that the problem can be directly mapped to the \textit{perfect hashing} problem. From this connection, similar bounds on the feedback rate can be found, which have the same dominant $\log(e)K$ scaling, plus a small $O(\log\log(N))$ term.

Furthermore, in certain applications, it is also of interest to consider the case where the available slots $B > K$, or the case where $B < K$ and up to $\left \lceil K/B \right \rceil$ users are permitted per slot and can be resolved subsequently via other means. 
In these cases, even fewer bits of feedback are required (see \cite{kang2020minimum}).

\begin{figure}[t]
\centering
\ifdefined\ONECOLUMN
\includegraphics[width=0.6\columnwidth]{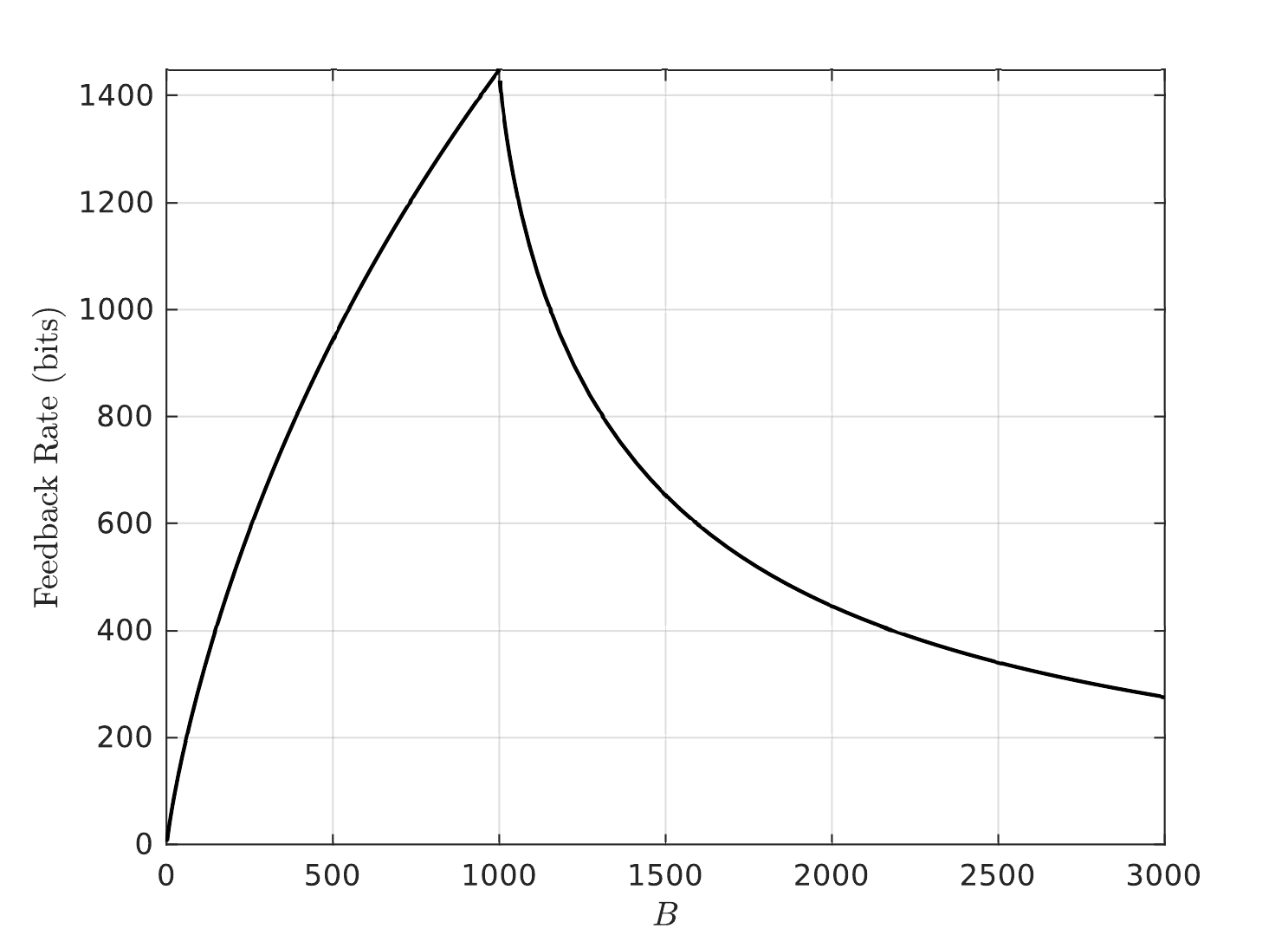}
\else
\includegraphics[width=0.9\columnwidth]{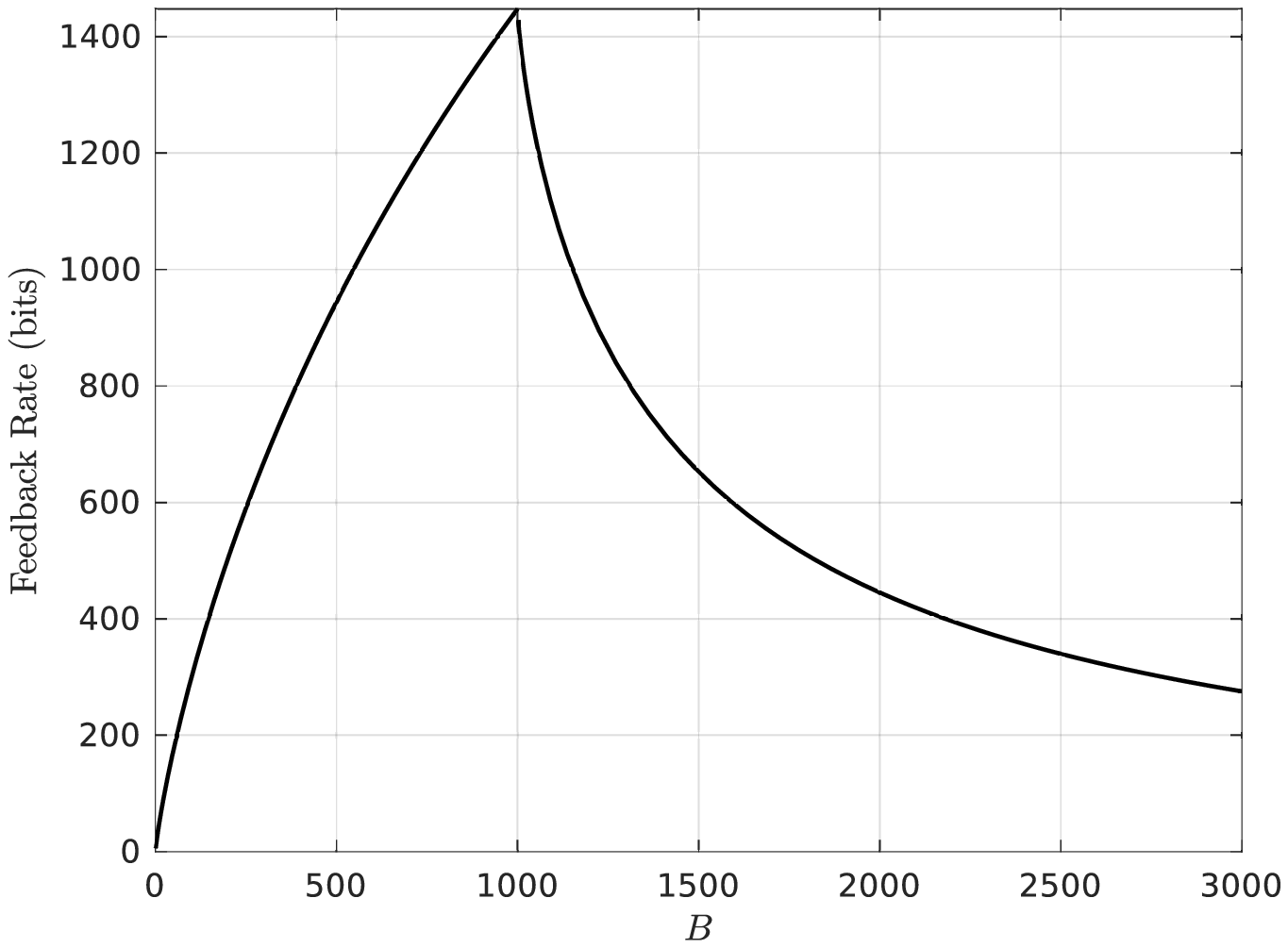}
\fi
\caption{The minimum feedback rate using variable-rate code for collision-free scheduling of $K=1000$ out of arbitrary number of $N$ users over $B$ slots.}
\label{fig:feedback}
\end{figure}

The above fundamental limit for collision-free feedback is significantly less than the naive scheme. For example, for $N=10^6$ and $K=B=1000$, the naive scheme would require $K \log(N) = 20000$ bits, while an optimal feedback code would only require at most $\log(e)(K+1) = 1444$ bits. 
Fig.~\ref{fig:feedback} plots the achievable bounds for the rate of variable-length feedback codes as function of $B$ for $K=1000$, which shows that if the number of slots is larger than $K$ or if multiple users can occupy the same slot, then the feedback rate can be significantly reduced. 
Exact expressions for these  bounds can be found in \cite{kang2020minimum}.
It can be seen from the figure that with just two users per slot, or with number of slots 15\% larger than the number of users, the minimum required feedback is already less than one bit per user!

\subsection{Feedback Scheduling for Massive MIMO Systems}

The possibility of highly efficient feedback is the main motive for this paper
to consider the benefit of scheduling for massive random access as compared to the
conventional contention based random access. 
Indeed, the aforementioned results show that the cost of scheduling can in theory be
as low as at most 1.44 bits per active user for the $B=K$ case.

The goal of scheduling for random access is to eventually separate users and to
avoid collision. Note that the above discussion has thus far referred to the concept of
``slots" in a deliberately abstract manner. For example, 
the notion of slot
does not need to be limited to simple temporal or frequency dimensions and can
also include the spatial or code domain. 

There are also situations in which multiple users can be scheduled into the same 
time-frequency slot, and they can be subsequently separated in the spatial domain
using beamforming or multiuser detection.
This scenario corresponds to the case of $B <K$.

Moreover, a key challenge of the massive MIMO system is in channel estimation.
Specifically, it is desirable to assign orthogonal pilots to the active
users in order to avoid pilot contamination in the channel estimation process.  
Note that the pre-assignment 
of orthogonal pilots to all potential users is not feasible, because there are 
too many potential users in the overall system and not enough orthogonal pilots. 

This paper proposes the use of efficient feedback codes for pilot assignment by
considering a correspondence between slots and unique orthogonal pilots. For 
massive MIMO systems in both the slow and fast-fading settings, we show that 
the use of a feedback strategy to assign unique orthogonal pilots to users for 
channel estimation can improve the system sum rate. 

\subsection{Practical Implementations}\label{sec:practical}

We now discuss the viability of practical implementation for the optimal collision-free feedback scheme. 
Fundamentally, the efficient feedback strategy amounts to constructing a list of hashing functions as the codebook, and for each given set of active users, searching for a hash function in the list that can map all the active users to distinct hashed values, then using the index of the hash function as the codeword. 

A practical implementation of such feedback code can be based on a procedure known as  the
\textit{compressed hash-displace} (CHD) method for perfect hashing \cite{BotelhoCHD2009}. 
The encoding procedure is effectively a two-level random hashing strategy of
first hashing users into bins, then starting from the bin with the most users,
hashing users into slots.  The random hash functions are drawn from an infinite
sequence of hash functions. A greedy strategy is used to search for the hashing
functions that result in no collision. The indices of the hashing functions,
properly compressed, is the feedback message.

In \cite{BotelhoCHD2009}, it is shown that this compression results in a code with a linear scaling in $K$.
The scaling coefficient depends on the choice of how many bins are used. Numerically it can be observed that having a larger number of bins makes the encoder faster, but at a cost of higher feedback rate. 
Indeed, when there is only one bin, the algorithm is exactly random hashing and requires an exponentially complex search over the sequence of hash functions but can achieve the $\log(e) K$ feedback rate. In practice, the choice of how many bins to use can be optimized to balance the trade-off between the complexity of encoding at the BS and the feedback rate.

Note that the feedback considered here is distinct from the traditional concept of a \textit{grant} \cite{Liu2018SparseThings} issued by the BS to the active user to acknowledge that its request for transmission has been received. The feedback scheduling codeword considered in this paper does not provide acknowledgement of detection to the active user and instead only serves to schedule the active users. If positive acknowledgement of detection is desired, the enumerative source coding scheme of \cite{Facenda2020} can be used, requiring $\log \binom{N}{K}$ bits of feedback. In many scenarios this is significantly more costly than the optimal feedback needed for avoiding collision. For a more detailed discussion of feedback for acknowledgement see \cite{Kalor2022}.
The lack of positive acknowledgement places fairly stringent requirement for user activity detection, because in the event of missed detection, the undetected active user 
would be unaware of the detection error and would transmit according to its decoded slot, leading to collision. For the falsely detected users, an allocated slot would be unoccupied, thus wasted. Fortunately, compressed sensing-based activity detection algorithms can operate at an error rate of $10^{-3}$ or less, thus alleviating these concerns.

\section{Scheduled Random Access in Fast-Fading Scenario} \label{sec:fast_fading}

In this section, we present the proposed three-phase random access scheme for the fast-fading channel model. 
An important point of reference for our proposed scheme is CPA.
CPA is a variant of CSA for massive MIMO \cite{Sorensen2018}. CPA operates by allowing users to contend for resources and potentially to collide with one another, but it then uses SIC to resolve the collisions. In contrast,  the proposed scheduled approach to massive random access exploits activity detection and feedback to enable scheduling and to prevent contention for resources between the users in the first place. Before presenting the scheduled approach random access, we present a brief summary of CPA to provide context for the discussion that follows.

\subsection{Coded Pilot Access}
CPA \cite{Sorensen2018} uses a simple repetition CSA scheme \cite{Casini2007} to add redundancy to
user transmissions and to enable collision resolution.  In this scheme each
active user in $\mathcal{A}$ transmits the same payload multiple times across 
multiple coherence blocks. In each coherence block $d$, whether a user transmits or 
not is based on the outcome of an independent Bernoulli trial with probability $p$ 
(where the value of $p$ can be optimized) such that each user transmits an average of $\beta = \Delta p$ times. Let the set of transmitting 
users in block $d$ be denoted as $\mathcal{A}_d$.
Inevitably, there would be collisions where two or more users transmit in the same
coherence block, i.e., $|\mathcal{A}_d| > 1$. In standard ALOHA, 
this would mean the loss of the payload and
the waste of a resource block. In CSA, however, the redundant transmissions may
allow the collisions to be resolved. For example, if the payload of one of the
users involved in the collision can be decoded in a different block where there
is no collision, the contribution from that user's transmission can be
subtracted from the collision. To resolve as many collisions as possible, a
graph based decoding scheme like those used for the erasure channel is used. 
 
\begin{figure}
\centering
\ifdefined\ONECOLUMN
\includegraphics[width=0.7\columnwidth]{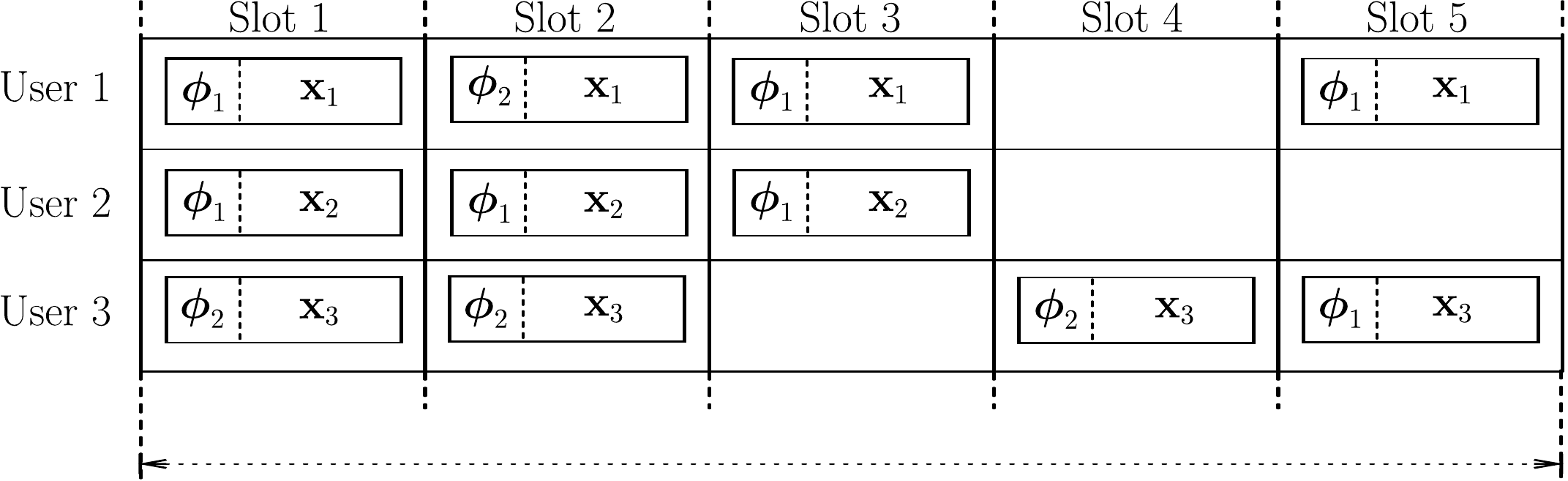}
\else
\includegraphics[width=0.9\columnwidth]{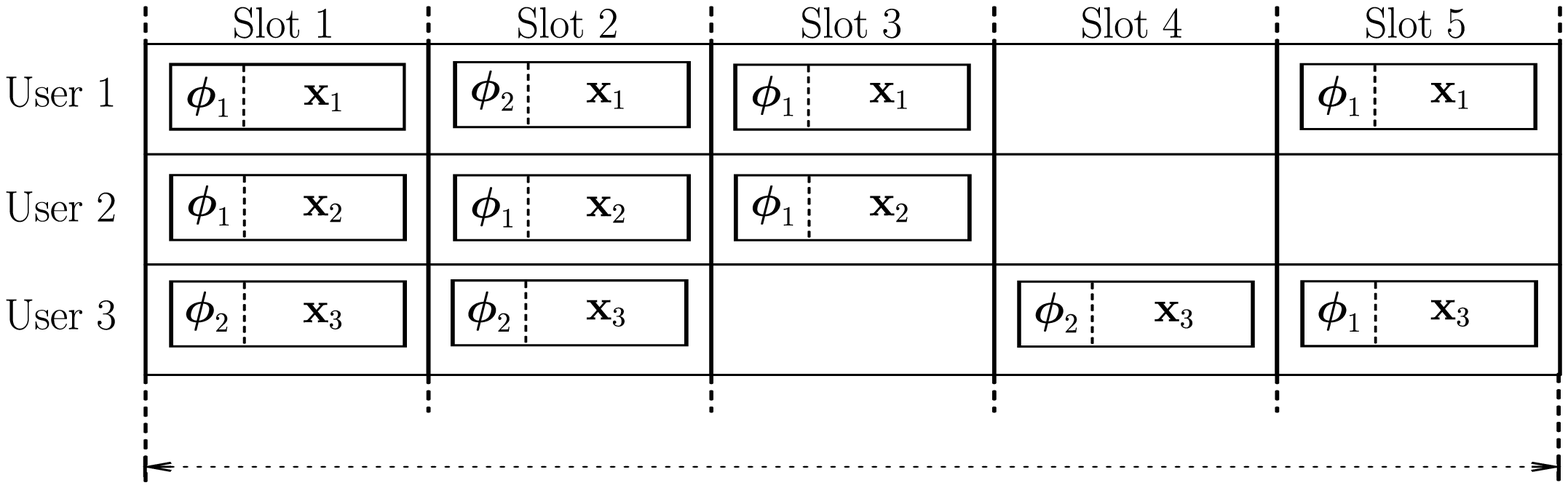}
\fi
\caption{Example of CPA with $K=3$, with $\tau=2$ orthogonal pilots, and a frame with $\Delta = 5$ coherence blocks.}
	\label{fig:cpa}
\end{figure}

CPA implements this coding scheme in massive MIMO by accounting for the need for channel estimation, and by providing a method for interference cancellation between coherence blocks. That is, even though user transmissions occur in a different coherence block where the channels are different, by incorporating a pilot, a payload decoded from one coherence block can still be used to subtract that user's contribution from the other blocks.

As shown in Fig.~\ref{fig:cpa}, in the CPA scheme,
the received signal $\mathbf{Y}_d \in \mathbb{C}^{M \times D}$ contains two portions
within each coherence block:
 $\mathbf{Y}^{(p)}_d \in \mathbb{C}^{M \times \tau}$, which is the pilot signals for channel estimation, and $\mathbf{Y}_d^{(u)} \in \mathbb{C}^{{M \times (D - \tau)}}$, which is the user payload.
In the pilot transmission phase of the coherence block, each active user $j \in\mathcal{A}_d$ selects a pilot $\boldsymbol{\phi}_{t_j} \in \mathbb{C}^{1 \times \tau}$ uniformly at random from the set of orthogonal pilots $\left\{ \boldsymbol{\phi}_i \right\}_{i=1}^{\tau}$. The signal received by the BS in this pilot phase can be written as:
\begin{equation}
\mathbf{Y}^{(p)}_d = \sum _{j \in \mathcal{A}_d} g_j\mathbf{h}_{d,j} \boldsymbol{\phi}_{t_j}  + \mathbf{Z}^{(p)}_{d}, \;\; d = 1,\dotsc,\Delta,
\end{equation}
where $\mathbf{Z}^{(p)}_{d}$ is an AWGN matrix.
At this stage, if only a single user has selected a given orthogonal pilot $\boldsymbol{\phi}_i$ in a coherence block, that user's channel may be estimated from the received signal.
If multiple users select the same $\boldsymbol{\phi}_i$ in the same block, however, a collision is declared. 

Following the transmission of pilots, users transmit their payloads:
\begin{equation}
\mathbf{Y}^{(u)}_d = \sum_{j \in \mathcal{A}_d} g_j\mathbf{h}_{d,j} \mathbf{x}_j + \mathbf{Z}^{(u)}_{d}, \;\; d = 1,\dotsc,\Delta.
\end{equation}
If a user is able to estimate its channel in a given coherence block, (i.e., it is
not involved in a collision), it can then attempt to decode its payload via 
receive beamforming. 
Each user's payload contains information about all the different blocks
throughout the frame where the user has made transmissions as well as the
identification information for that user. 
Even though the user's instantaneous CSI would differ
between transmissions in different blocks, \cite{Sorensen2018} shows that by
exploiting the properties of massive MIMO, specifically channel hardening and
the temporal power stability of the instantaneous channel,  once a payload has
been decoded, the associated user's contribution to the received signal in
other blocks can be subtracted. Decoding then proceeds in the graph as in 
erasure decoding.

Although CPA can outperform traditional SA based random access schemes, it still may not be able to fully utilize all the coherence blocks. For example, the way the users choose slots for transmission in an i.i.d.\ fashion as in \cite{Sorensen2018} induces a binomial degree distribution on the user nodes in the decoding graph. This is suboptimal even with an optimized Bernoulli trial probability $p$. It is well known that for erasure decoding, the \textit{soliton} distribution \cite{Narayanan2012} is an asymptotically optimal degree distribution, achieving perfect utilization as the number of blocks and number of users go to infinity. But as the number of coherence blocks in a frame is typically small in practical systems, this asymptotic performance cannot typically be achieved. Finally, as in all SA schemes, the fact that each user transmits in multiple coherence blocks results in unnecessary additional energy consumption.

\subsection{Scheduled Random Access}
\begin{figure}[t]
\centering
\ifdefined\ONECOLUMN
\includegraphics[width =0.55 \columnwidth]{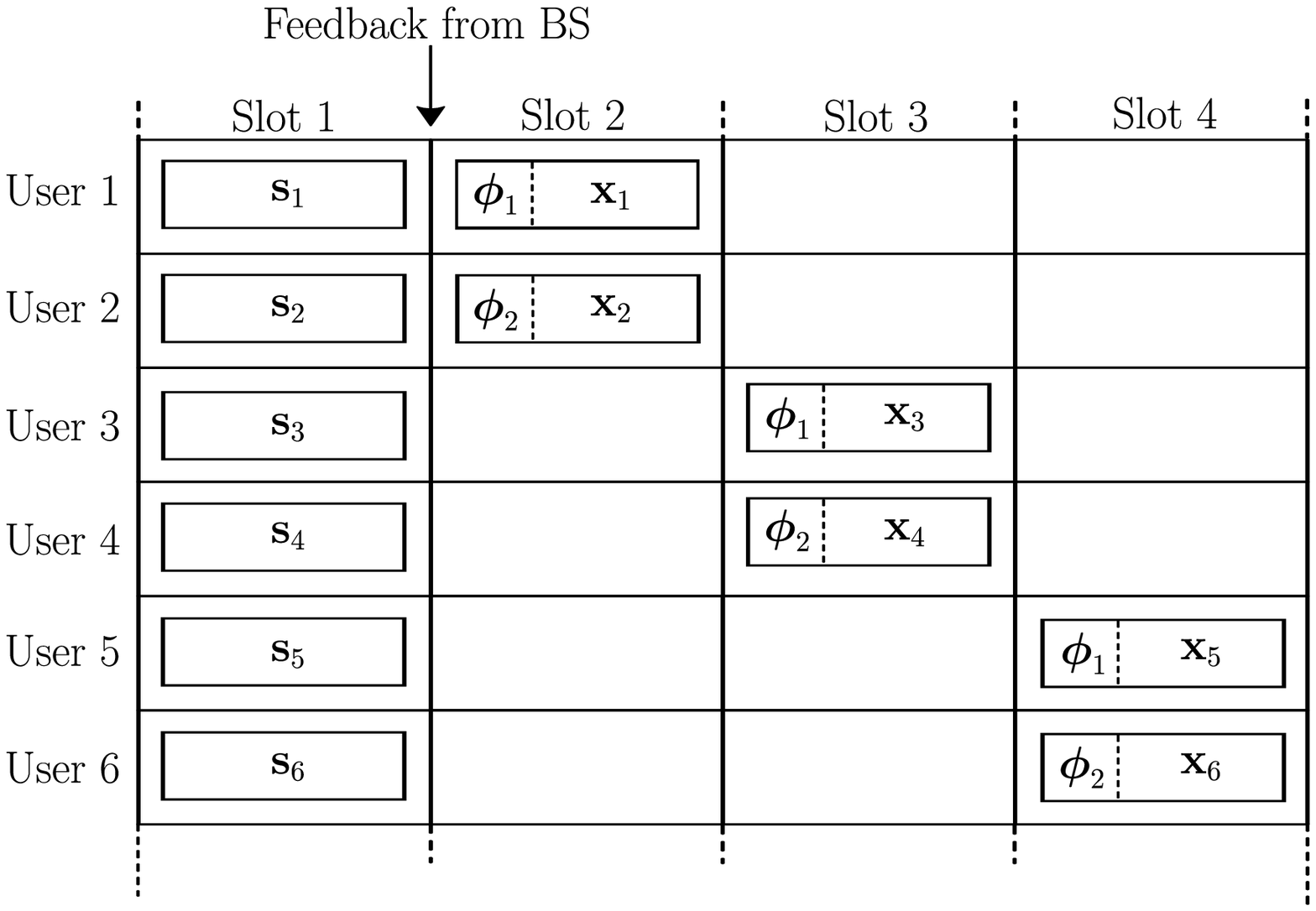}
\else
\includegraphics[width =0.9 \columnwidth]{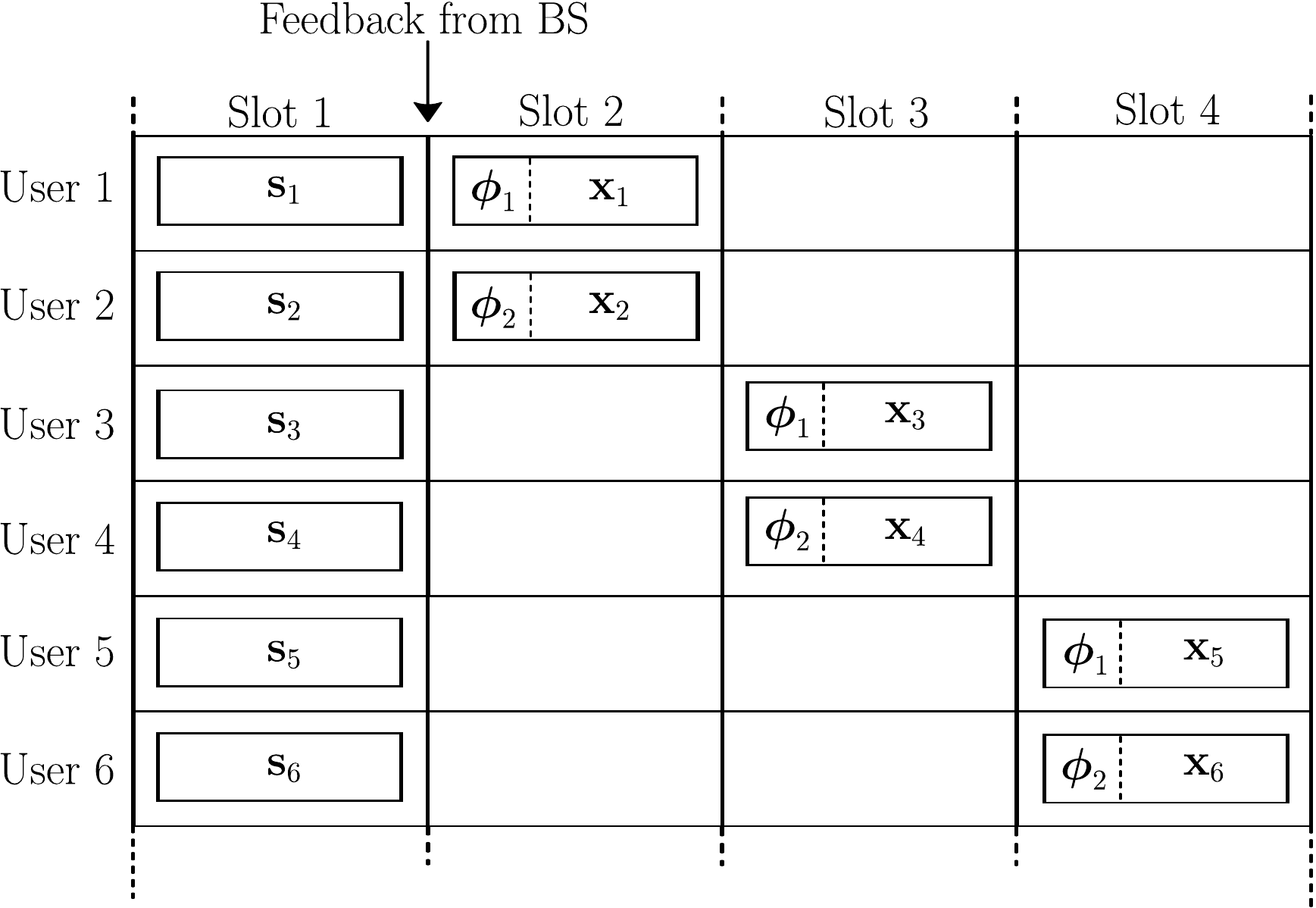}
\fi
\caption{Three-phase coordinated random access scheme for the fast-fading scenario. Activity detection occurs in the first phase where the $K$ active user transmit non-orthogonal pilot sequences. The BS detects the active users, then transmits a feedback message to schedule them into the remaining coherence blocks along with orthogonal pilots within each block.}
\label{fig:scheduled}
\end{figure}

Scheduled random access offers an alternative to the contention-based CPA. Rather than using redundancy to resolve contention, the proposed scheduled approach instead allocates an initial block for activity detection, followed by a short feedback message from the BS to the users for scheduling. Then, all remaining blocks are perfectly utilized, with no slots wasted for unresolved collisions. Fig.~\ref{fig:scheduled} depicts this three-phase procedure for the fast-fading channel model. 
This scheme offers multiple advantages over  CPA. First, in the scheduled approach, all slots are perfectly utilized, with the only overhead being the initial activity detection phase and the feedback, while in CPA, perfect utilization cannot be achieved, and the overhead is more significant. Additionally, this scheduled approach only requires the users to transmit twice: once to transmit non-orthogonal pilots for activity detection and once for data transmission. In contrast, to maximize throughput with CPA, the users often must transmit more than twice, resulting in excess power consumption, which is a critical issue for IoT applications.

Below we describe the proposed scheme in more detail.

\subsubsection{Activity Detection}\label{sec:act_detect}
The first coherence block is dedicated to activity detection. In this block, all active users simultaneously transmit pre-assigned non-orthogonal pilot sequences.
We use the covariance approach for activity detection. 
This is because as mentioned in Section \ref{sec:activity}, in the fast-fading model, CSI changes between the activity detection and data transmission
phases, thus the CSI estimate provided by the AMP algorithm is not useful in subsequent blocks. Further, due to the complexity
scaling of AMP with respect to the number of antennas $M$, the covariance approach offers a lower complexity and generally superior detection performance.

It may be the case, depending on system parameters, that one coherence block is insufficient for accurate activity detection. For example, if the SNR is too low, a single coherence block may not allow the active users to transmit pilot sequences long enough to enable accurate activity detection. In such settings, more than one coherence block in the frame can be dedicated to activity detection, at the cost of decreasing the number of remaining slots for payload transmission. 
In this paper, the system parameters are chosen such that 
a single coherence block is sufficient for activity detection, as shown in the next section.

\subsubsection{Scheduled Transmission}

Collision occurs when 
two or more users select the same coherence block for transmission \textit{and} the same pilot for channel estimation. Thus, in scheduled transmission the BS needs to use feedback to allocate the transmission blocks as well as the pilots for all the active users. 
To this end, consider a pool of $\tau$ orthogonal pilots $\boldsymbol{\phi}_1, \boldsymbol{\phi}_2, \dotsc, \boldsymbol{\phi}_{\tau} \in \mathbb{C}^{\tau}$ for channel estimation in each coherence block. We use feedback from the BS to assign each active user a unique block-pilot pair $(d,t) \in \left[ 2, \Delta \right] \times \left[ \tau \right]$. Thus, the number of effective slots for the purpose of scheduling is $B = \tau(\Delta - 1)$. To avoid collision, we  must have $B \geq K$, so $\tau$ should be chosen accordingly.
Since each user requires a unique pair, the minimum amount of feedback required is approximately $\log(e) K$ bits. Note that up to $\tau$ users can transmit in the same coherence block, as long as they are assigned distinct orthogonal pilots, so that their channels can be properly estimated. This allows their payloads to be resolved spatially in the massive MIMO system. 

Let $(d_i, t_i)$, $i \in \mathcal{A}$ be the block-pilot pair decoded  by the active users based on the feedback from the BS. Let $\mathcal{A}_d \triangleq \left\{ i\; |\; d_i = d\right\}$ be the set of active users that have been scheduled to transmit in the coherence block $d$. 
Just as in CPA, each user's data transmission within a block is split into two parts. Let $\mathbf{Y}^{(p)}_d \in \mathbb{C}^{M \times \tau}$ denote the signal received by the BS, we have
\begin{equation}
\mathbf{Y}^{(p)}_d = \sum _{j \in \mathcal{A}_d} g_j \mathbf{h}_{d,j} \boldsymbol{\phi}_{t_j}  + \mathbf{Z}^{(p)}_{d}, \;\; d = 2,\dotsc,\Delta.
\end{equation} 
Immediately after transmitting a pilot, each user in $\mathcal{A}_d$ transmits their payloads simultaneously. Let $\mathbf{Y}_d^{(u)} \in \mathbb{C}^{{M \times (D - \tau)}}$ denote the signal received by the BS in this stage, we have
\begin{equation}
\mathbf{Y}^{(u)}_d = \sum_{j \in \mathcal{A}_d} g_j \mathbf{h}_{d,j} \mathbf{x}_j + \mathbf{Z}^{(u)}_{d}, \;\; d = 2,\dotsc,\Delta.
\end{equation}
If activity detection is perfect, for each user $i \in \mathcal{A}$ the BS can obtain a channel estimate $\hat{\mathbf{h}}_{i,d_i}$ based on $\mathbf{Y}^{(p)}_{d_i}$ (e.g., via a least-squares estimator \cite{Sorensen2018}).
Finally, the estimated channel is used to separate the transmitted signals in $\mathbf{Y}^{(u)}_{d}$ via receive beamforming so that all the payloads can be decoded. 

\subsection{Performance Evaluation}

In this section we numerically compare the performance of the proposed scheduled random access scheme with CPA. We follow \cite{Sorensen2018} by assuming that the users apply inverse power control so $g_i = 1$, and further assume 
that the SNR is $10$dB. In addition, the base station is equipped with $M=400$ antennas and employs maximum ratio combining. Furthermore, there are $N=10000$ potential users. We assume a channel bandwidth and coherence time of $1$MHz and $0.3$ms respectively, meaning $D=300$ symbols can be transmitted in each coherence block.  We assume a latency constraint such that $\Delta = 15$. We fix the number of orthogonal pilots in the frame to be $\tau=64$, where each pilot consists of $\tau$ symbols.
Furthermore, to simplify simulation we assume perfect interference cancellation in CPA, so the performance plot serves as an upper bound for CPA. 

For scheduled random access, simulation results indicate that with activity detection pilots of length $L = 300$ and $K=1000$, the user activity detector based on the covariance approach can achieve a probability of false alarm $p_{\rm FA} = 10^{-3}$ and a probability of missed detection $p_{\rm MD} = 10^{-4}$. Thus, a single coherence block provides sufficiently accurate activity detection in this setting.  

\begin{figure}[t]
\centering
\ifdefined\ONECOLUMN
\includegraphics[width=0.7\columnwidth]{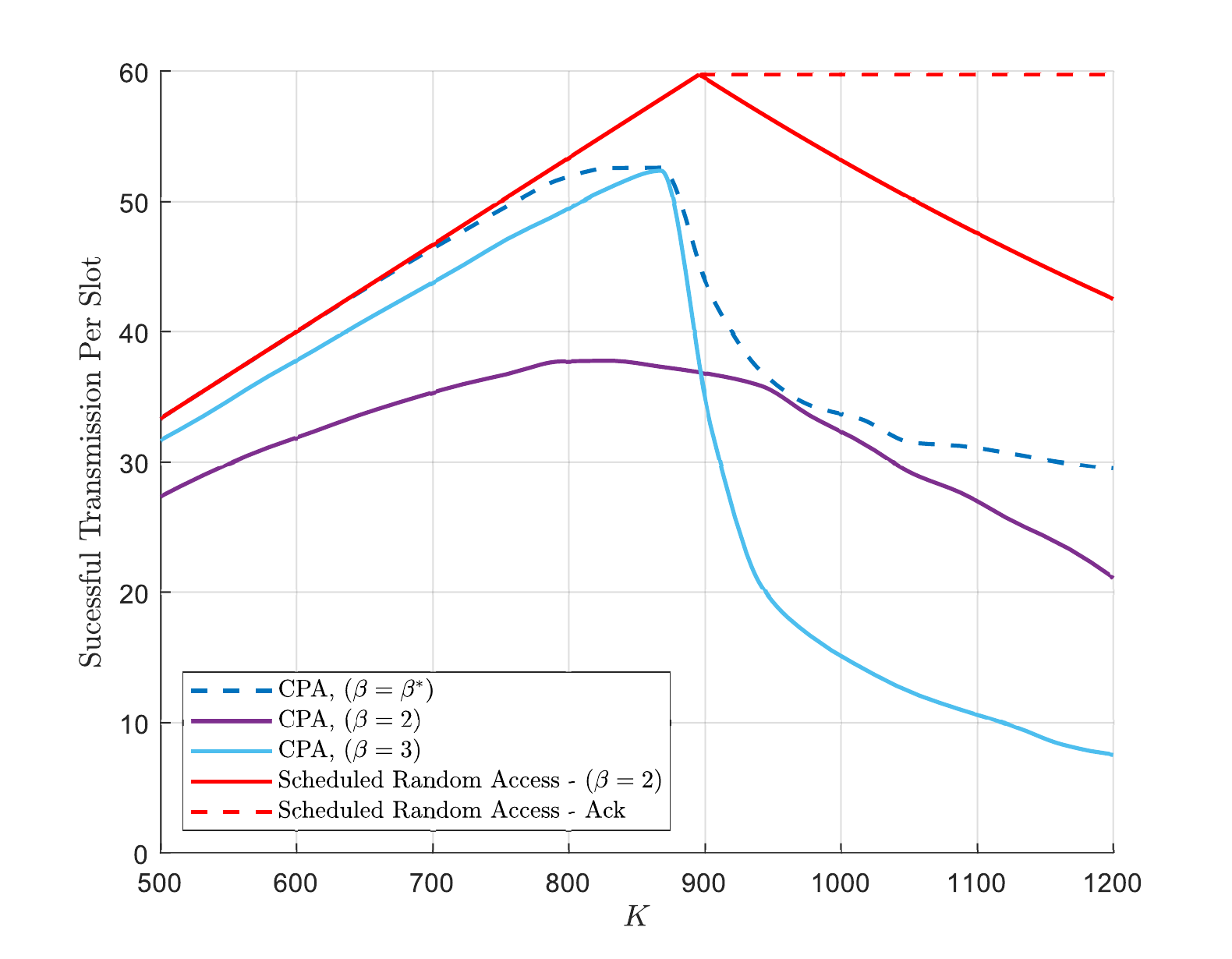}
\else
\includegraphics[width=\columnwidth]{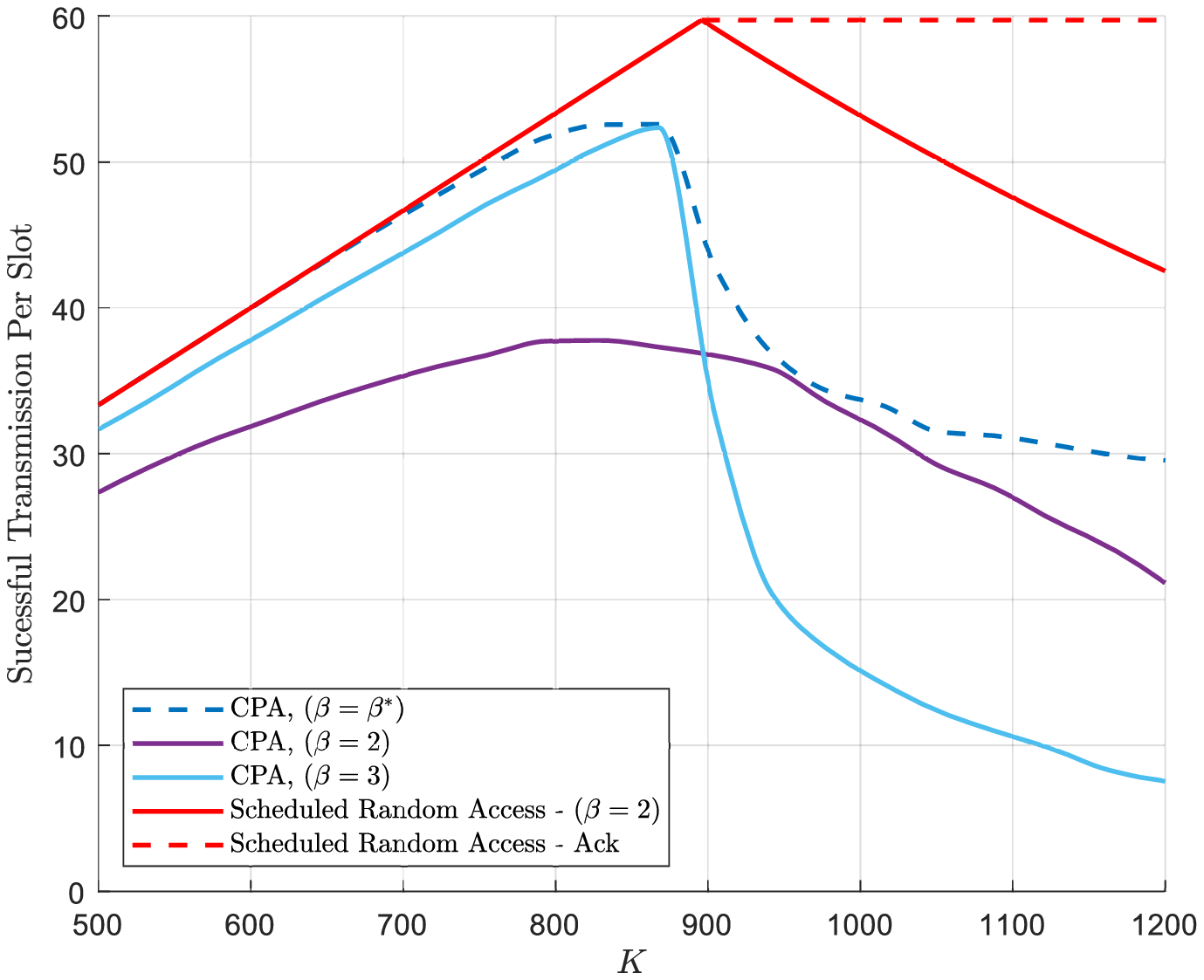}
\fi
\caption{The number of successful transmissions per slot for scheduled random access and CPA at $\Delta=15$, $N=10000$, $M=400$, $L=300$ and $\tau=64$ for a range of values $K$.}
\label{fig:cpa_vs_spa}
\end{figure}

Fig.~\ref{fig:cpa_vs_spa} compares the performance of CPA and scheduled random access under these settings. 
We use the number of \emph{successful transmissions} per slot as the performance metric. For scheduled random access, this is the number of active users that are detected, scheduled, and transmit without collision, divided by $\Delta$. For CPA, this is the number of ``singleton" users that are successfully decoded, divided by $\Delta$. 
Note that the performance of CPA depends crucially on the average number of transmissions per user, which is a parameter denoted as $\beta$. 

For scheduled random access, we assume that a user that transmits a pilot not used by any other user in the same slot would always be able to successfully transmit their payload. This effectively assumes perfect channel coding. In CPA, we make a similar assumption, and further assume that SIC can be done perfectly. Note that these assumptions favor CPA, because a payload suffering a decoding failure due to insufficient channel coding only impacts that particular payload in the scheduled random access, while in CPA it can prevent SIC and impact the ability to decode other payloads as well. 

The interpretations of the simulation results presented in Fig.~\ref{fig:cpa_vs_spa} are as follows:

\subsubsection{Scheduled Random Access}
Since we operate in the regime where activity detection is accurate, the impact of detection error is negligible. Thus, so long as the number of users is less than the maximum number of available slots, i.e., $K \le \tau(\Delta-1)$, all users can be accommodated, and the number of successful transmissions grows linearly in $K$ in this regime as can be seen in Fig.~\ref{fig:cpa_vs_spa}. If $K > \tau(\Delta-1)$ there are more users than can be scheduled. 
If the feedback message does not provide positive acknowledgement (e.g., as in a CHD-based feedback coding scheme \cite{kang2020minimum} as discussed in Section~\ref{sec:practical}), then the BS cannot prevent active users from transmitting. 
In this case, despite that the BS schedules only a subset of detected users of size $\tau(\Delta-1)$, due to the collision from the active users not explicitly scheduled, the number of successful transmissions would decrease in the regime $K > \tau(\Delta-1)$ as plotted in Fig.~\ref{fig:cpa_vs_spa}. 
However, if positive acknowledgment is provided (e.g., as in \cite{Kalor2022}) at the cost of additional feedback bits, then the BS can use feedback to ensure that only the maximum number of supported users would transmit, resulting in no drop-off. 
Note that the scheduled random access performs best when $K = \tau(\Delta-1)$, however, as in general the number of active users may vary from frame to frame based on the arrival process, it is not possible to always operate at this setting.

The amount of feedback required to achieve this performance 
is given in \cite[Section V]{kang2020minimum} (e.g., Fig.~\ref{fig:feedback}
for $B=1000$). This scheme requires $B$ to be known prior to the start of the frame, which is also a requirement in CPA. For the setting of Fig.~\ref{fig:cpa_vs_spa}, the amount of feedback is at most $1.3$ kbits, which is a small fraction (typically $<1\%$) of the overall throughput. 

\subsubsection{CPA}
Fig.~\ref{fig:cpa_vs_spa} presents several plots depicting the performance of CPA. The dashed curve presents the number of successful transmissions per slot after optimizing over $\beta$, the average number of transmissions per user. The dashed curve increases with $K$ until $K \approx 0.9 \tau \Delta$, at which point performance begins to sharply drop off, showing that the approach requires additional slots to be effective, and does not perform well if too many users attempt access. This dashed curve serves as an upper bound on the performance of CPA and is generally unattainable without an accurate estimation of $K$, as the optimal $\beta^*$ in general depends on $K$. 

We also plot the performance of CPA at different power levels. When $\beta = 2$, each user transmits twice, and the average uplink power requirement matches that of scheduled random access. But for most values of $K$, $\beta = 2$ is less than the optimal value $\beta^*$, and the performance degradation is significant. We also present results for $\beta = 3$, which is closer to optimal. For $K \lessapprox 880$, $\beta^* > 3$ and for $K 	\gtrapprox 880$, $\beta^* < 3$.

\subsubsection{Comparison} 
Fig.~\ref{fig:cpa_vs_spa} shows that when there are many more available slots than active users, CPA with $\beta^*$ and scheduled random access perform similarly, but optimal CPA requires on average up to $50\%$ more power in the uplink, while scheduled random access requires feedback from the BS to the users. Furthermore, decreasing $\beta$ in this regime leads to significant degradation in the performance of CPA.

As the number of users approaches the maximum number of available slots, however, the performance gain of the scheduled approach over CPA becomes apparent. The performance of CPA begins to diminish rapidly around $K \approx 0.9 \tau \Delta$, while the performance of scheduled random access is significantly more stable, peaking at $K = \tau(\Delta - 1)$. 

These results show that when operating in a regime where $K$ is close to the
maximum, scheduled random access has a considerable advantage. 
For CPA to maximize the number of successful transmissions per slot, it requires 
an additional overhead of some fraction of slot-pilot pairs. In general, 
we may wish to operate at $(K, \tau\Delta)$ for which
the number of successful
transmissions per slot is maximized. However, in practice, user activities 
vary with time depending on the arrival process (e.g., as modeled by a
Poisson or Beta arrival model) with inherent randomness in $K$.  Thus, 
a random access protocol must perform well not only for one choice of $K$, 
but also over a range of potential values of $K$. In other words, the stability 
of the random access scheme around its optimal operating point is important,
so that the system can be loaded more aggressively. This is a consideration 
that would favor the proposed scheduled approach as compared to CPA.

Note that the above analysis does not account for the extra overhead in the payload of CPA, which requires each user to transmit identification information, as well as pointers needed for interference cancellation. Although the cost of pointers can be considered negligible, the cost of identification is $\log(N) \approx 13$ bits per user, 
which must be included in the payload in CPA, but not in the proposed scheduled approach. 
In conclusion, a small amount of feedback can significantly improve the overall number of successful transmissions per slot in scheduled random access, while using less power per user as compared to contention-based schemes for a massive MIMO system in the fast-fading scenario.

\section{Scheduled Random Access in Slow-Fading Scenario}\label{sec:slow_fading}

In this section we investigate the case of the slow-fading channel model, where
all communications occur over a single coherence block. This situation occurs
when the CSI changes slowly relative to the time scale of user activities. 
For example, the system may have a latency requirement dictating that
an active user must be served in a period of time shorter than the coherence
block.  As compared to the
fast-fading case, the main difference in slow fading is that each user's
instantaneous CSI is the same in both the activity detection and data
transmission phases. This means that any pilot symbols initially used for
activity detection can subsequently be re-used for channel estimation. In this
section, we discuss two protocols.  One uses non-orthogonal pilots to estimate
the user channels, while the second scheme uses feedback to assign orthogonal
pilots to different users for channel estimation. 

\subsection{Joint Activity Detection and Channel Estimation Using Non-Orthogonal Pilots}

\begin{figure}[t]
\centering
\ifdefined\ONECOLUMN
\includegraphics[width=0.6\columnwidth]{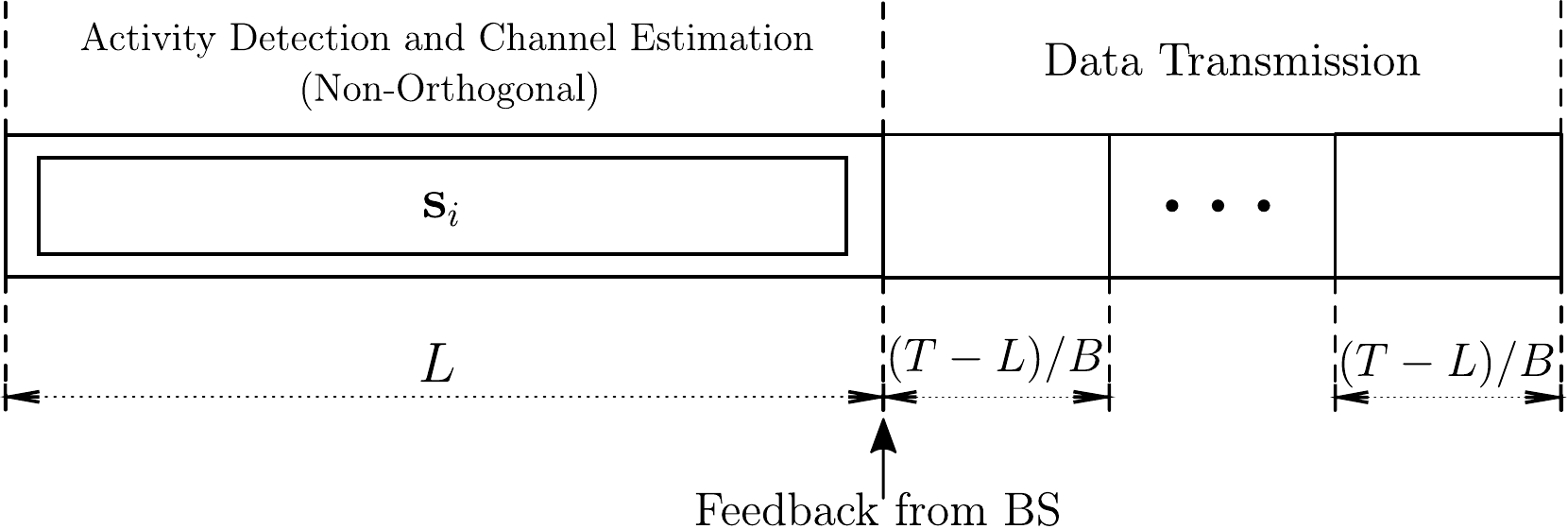}
\else
\includegraphics[width=0.9\columnwidth]{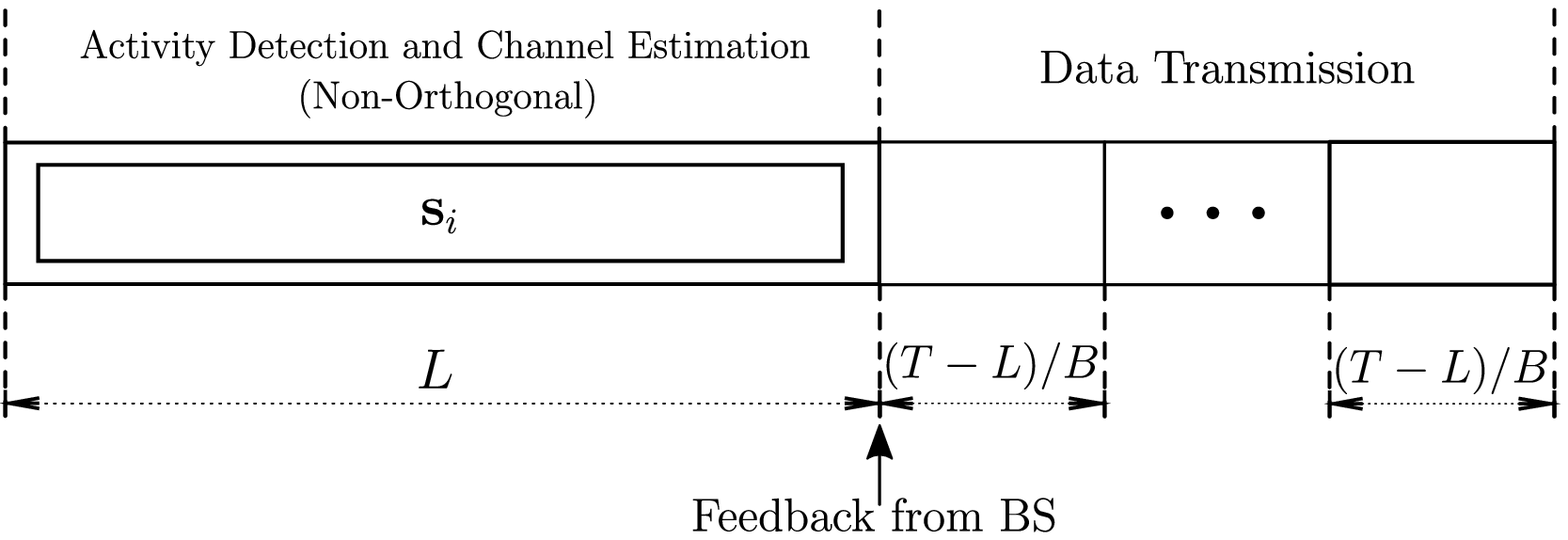}
\fi
\caption{Slow-fading scenario with joint user activity detection and channel estimation using non-orthogonal pilots. }
\label{fig:slow_fading_amp}
\end{figure}

The paradigm of \textit{joint activity detection and channel estimation} is investigated in \cite{Liu2018_1, Liu2018}. In that work, a frame of length $T$ symbols is separated into two phases, with $L$ symbols for joint activity detection and channel estimation, and the remaining $T-L$ symbols for data transmission. The activity detection phase remains identical as in Section \ref{sec:activity}, but additionally, because the problem is formulated as an MMV compressed sensing problem \eqref{eq:AMP_mmv}, the estimated value of the CSI matrix $\mathbf{X}$, can be useful for receiver design in the subsequent data transmission phase. This scheme does not require feedback, because in
the data transmission phase all users can transmit simultaneously and the BS can separate their signals using receive beamforming based on the previously estimated CSI. 

One of the problems with this approach is that when too many users are transmitting simultaneously, interference can lead to a significant degradation in the overall rate. Specifically, when the system is \textit{overloaded}, meaning $K/M > 1$, it can be advantageous to temporally schedule users, i.e., the transmission phase is divided into $B$ non-overlapping slots of $\frac{T-L}{B}$ symbols, such that in each slot only at most $\left \lceil K/B \right \rceil$ users are scheduled for transmission.  The spectral efficiency for each user averaged over the frame can be written as:
\begin{equation}
R^{NO}_{k} = \frac{T-L}{TB} \log(1+ \gamma^{NO}_k),
\end{equation}
where $\gamma^{NO}_k$ is the signal-to-interference-and-noise-ratio (SINR), which includes the interference terms due to the users who transmit in the same slot as well as the effect of the channel estimation error and the AWGN. The asymptotic value of $\gamma^{NO}_k$ can be computed via the state evolution of AMP. It can be shown that when MMSE receive beamforming is used, scheduling can improve the achievable sum rate of an overloaded system \cite{Liu2018}.
 This means that we can optimize the system over $B$ to determine the optimal $B^*$ that maximizes the user sum rate. 

Although this scheme is already investigated in \cite{Liu2018}, the minimum cost of scheduling and feedback has not been previously quantified. In this work, we note that this is a case of scheduling multiple users per slot as investigated in \cite{kang2020minimum}, so the information-theoretic bounds from \cite{kang2020minimum} can be used to determine the amount of feedback required for scheduling. Since the BS is equipped with many antennas, in most cases we have $B^* \ll K$. In this regime, Fig~\ref{fig:feedback} shows that significantly less than $K \log(e)$ bits of feedback are needed. 
Note that when $B=1$, this approach reduces to the previous no feedback case.

\subsection{Scheduling of Orthogonal Pilots for Channel Estimation}

Despite the promise of using non-orthogonal pilots for both activity detection and channel estimation, one of the key conclusions of \cite{Liu2018} is that the bottleneck in performing joint activity detection and channel estimation lies in the use of non-orthogonal pilots for channel estimation, which leads to a significantly larger channel estimation error as compared to if orthogonal pilots are used, resulting in lower achievable rates.

We propose to resolve this issue via a natural extension to the previously discussed scheduling strategy. Since effective activity detection for $K$ users requires shorter pilots than channel estimation for those same $K$ users, we propose to perform activity detection using non-orthogonal pilots of length $L_1$ via the covariance approach, then subsequently to provide a feedback message as in Section \ref{sec:feedback} to assign orthogonal pilots of length $L_2$ to each of the active users for a second channel estimation phase. 
Here the covariance approach is suitable for the first phase, because the BS does not require estimates of the channels at the intermediate stage after the non-orthogonal pilot transmissions. 
After the orthogonal pilots are transmitted by the active users, the BS can use both the non-orthogonal pilots in the first phase and the orthogonal pilots in the second phase to perform channel estimation using a linear MMSE channel estimator \cite{Biguesh2004}. 
Explicitly, we can write the estimate 
of the channels between the $K$ active users and the $M$ BS antennas ${\mathbf{H}}^T_{ \rm MMSE} \in \mathbb{C}^{K \times M}$ as
\begin{equation}
\hat{\mathbf{H}}^T_{ \rm MMSE} = \mathbf{Y}\left( \mathbf{P}^H \mathbf{R}_{\mathbf{H}} \mathbf{P} + \sigma^2_n \mathbf{I}\right)^{-1}\mathbf{P}^{H}\mathbf{R_{H}},
\label{eq:MMSE}
\end{equation}
where $\mathbf{Y} \in \mathbb{C}^{M \times L}$ is the received signal during the
channel estimation phase, $\mathbf{R}_{\mathbf{H}} =
\mathbb{E}[\mathbf{H}\mathbf{H}^H]$ is the channel correlation matrix,
and $\mathbf{P} \in \mathbb{C}^{K \times L}$ is a matrix that has rows equal to
the pilots of the $K$ active users. These pilots can be either non-orthogonal,
or a concatenation of orthogonal and non-orthogonal pilots transmitted by the
active users.
 
\begin{figure}[t]
\centering
\ifdefined\ONECOLUMN
\includegraphics[width=0.6\columnwidth]{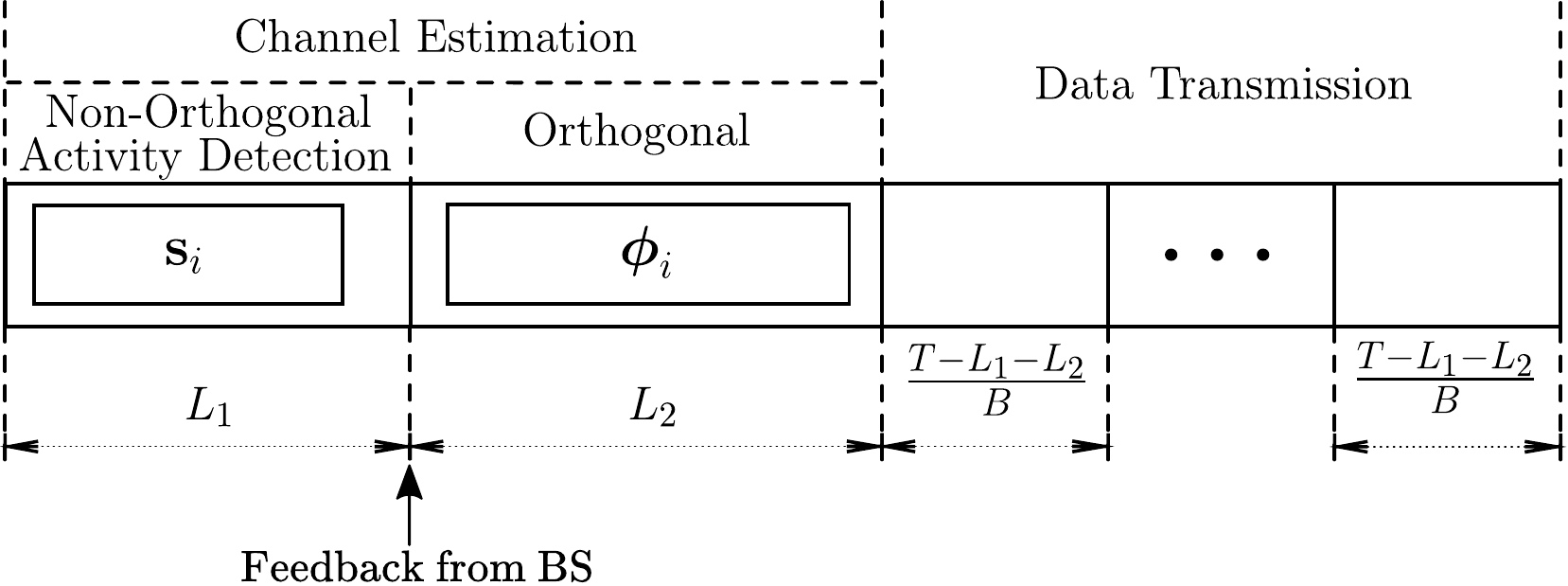}
\else
\includegraphics[width=0.9\columnwidth]{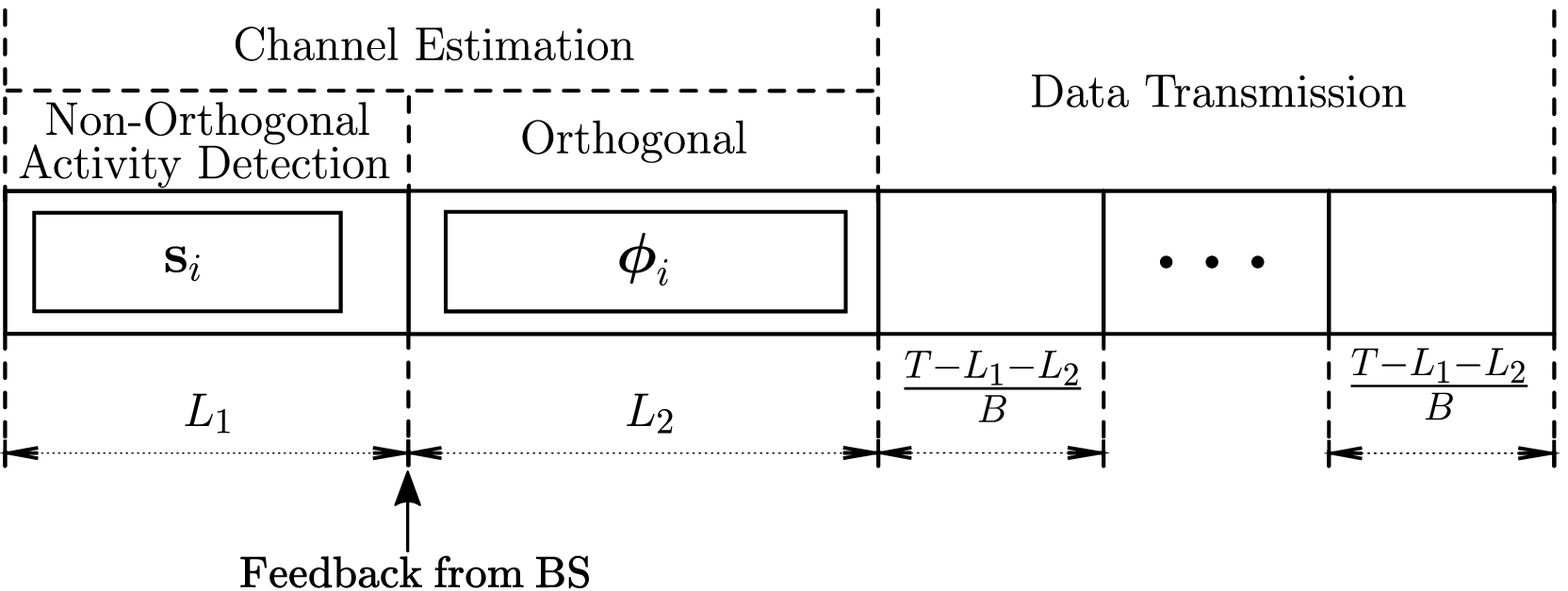}
\fi
\caption{Slow-fading scenario with channel estimation based on orthogonal pilots.}
\label{fig:slow_fading_cov}
\end{figure}

The use of orthogonal pilots eliminates pilot contamination between the users, resulting in an overall sum rate gain due to the improved channel estimates.
In addition, the same feedback message used to assign pilots can be re-used to schedule users across $B$ slots, and the optimal $B^*$ can be chosen to maximize the user sum rate. 
The overall scheme is shown in Fig \ref{fig:slow_fading_cov}.

The per-user spectral efficiency averaged over the frame for the proposed scheme can be characterized as:
\begin{equation}
R^O_{k} = \frac{T-L_1 - L_2}{TB} \log(1+ \gamma^O_k)
\end{equation}
where $\gamma^O_k$ is the SINR which includes the effect of the channel estimation error when both non-orthogonal and orthogonal pilots are used. As with the previous case, MMSE receive beamforming is used for data transmission. To compute the SINR, we follow the method in \cite[Appendix A] {Shen2019} but with the MMSE beamforming. The method accounts for the choice of the pilots, the resultant channel estimation error, as well as the distribution of the channels, in computing the SINR. 

The proposed scheme assigns an orthogonal pilot to each user, so it requires more feedback than the case of using only non-orthogonal pilots. Since the pilot indices are unique, at least $K\log(e)$ bits of feedback would be required. No additional feedback is needed to schedule users into the $B$ slots as the pilot index can be re-used to determine the slot. For example, a user who is assigned pilot index $k$ can be assigned to transmit in slot $b = k\; \text{mod}\; B$. All the users are then distributed uniformly over the $B$ slots.

\subsection{Performance Evaluation}

To evaluate the benefit of assigning orthogonal pilots for channel estimation in the slow-fading scenario,
we consider a simulation setup like that in \cite{Liu2018}. Let $d_n$ denote the distance between user $n$ and the BS, $\forall n$. It is assumed that each $d_n$ is randomly distributed in the range $[0.8, 1]$ km. The path-loss model of the wireless channel for user $n$ is given as $ -128.1 - 36.7\log_{10}{d_{n}}$ in dB $\forall n$. The bandwidth and coherence time of the channel are $1$MHz and $2$ms respectively, thus there are a total of $T=2000$ symbols per frame. The transmit power is constant across the coherence block and is set to be $13$dBm. The power spectral density of the AWGN at the BS is $-169$dBm/Hz.

First, we investigate the impact of the number of temporal slots $B$ used for scheduling data transmission for both the case that non-orthogonal pilots are used and the case that orthogonal pilots are used for channel estimation. The optimal $B$ is a function of the frame length $T$, the pilot length $L$, and the other system parameters. As an example, consider a scenario with $M=64$ antennas at the BS and $K=150$ of the $N=2000$ users are active. (This smaller value of $M$ is chosen so that we operate in a regime where the complexity of the AMP algorithm remains tractable.) 
In the case that orthogonal pilots are assigned for channel estimation, we first use non-orthogonal pilots of length $L_1=200$ for activity detection, then we use feedback to assign orthogonal pilots of length $L_2 = L - L_1$ to the active users for channel estimation. With $L_1 = 200$ and under these conditions, simulation results show that if the covariance approach is used for activity detection, the probability of false alarm is $p_{\rm FA} < 10^{-5}$ and the probability of missed detection is $p_{\rm MD} = 10^{-4}$. 
This shows that $L_1=200$ is adequate.

\begin{figure}[t]
\centering
\ifdefined\ONECOLUMN
\includegraphics[width=0.6\columnwidth]{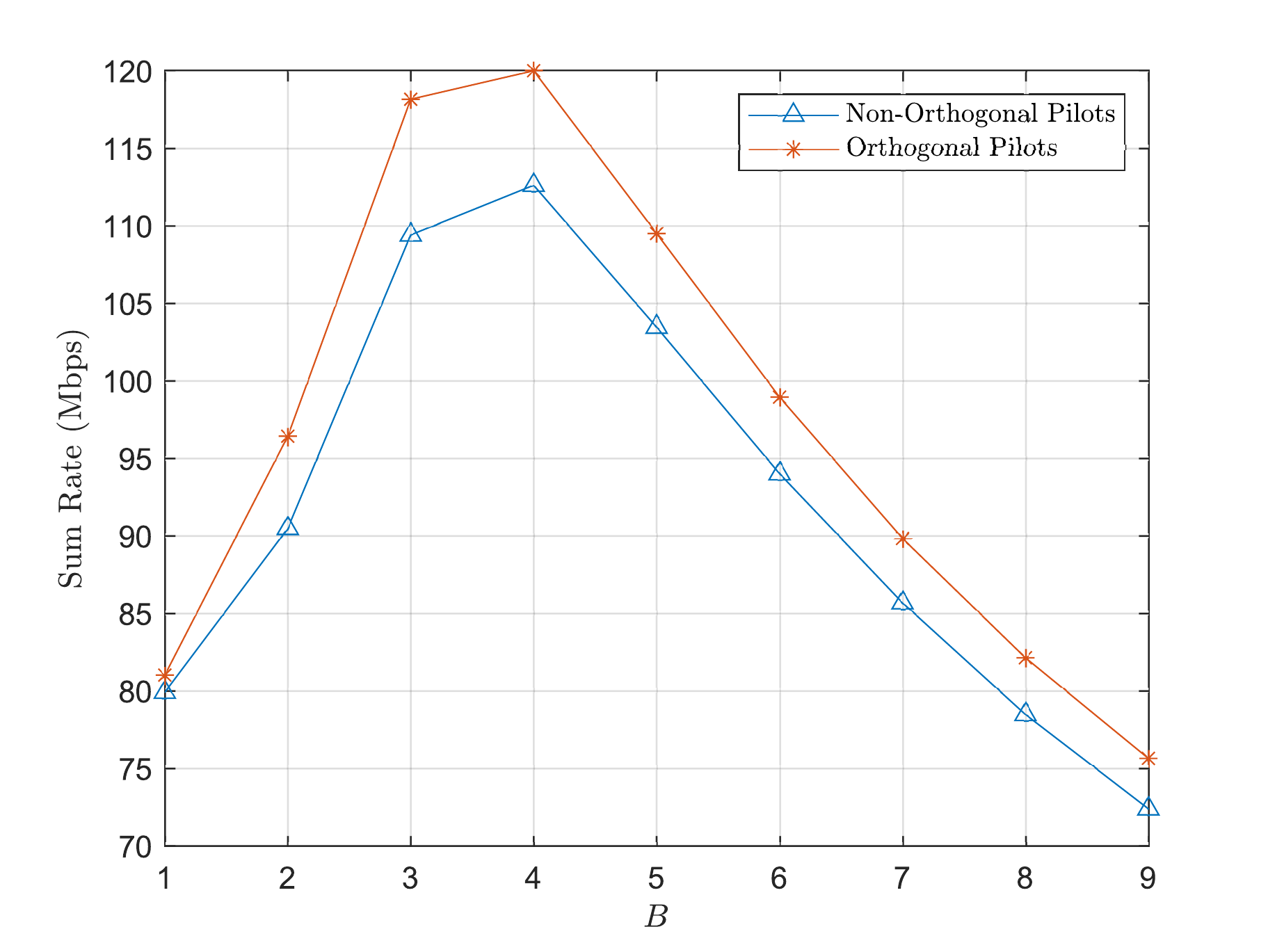}
\else
\includegraphics[width=\columnwidth]{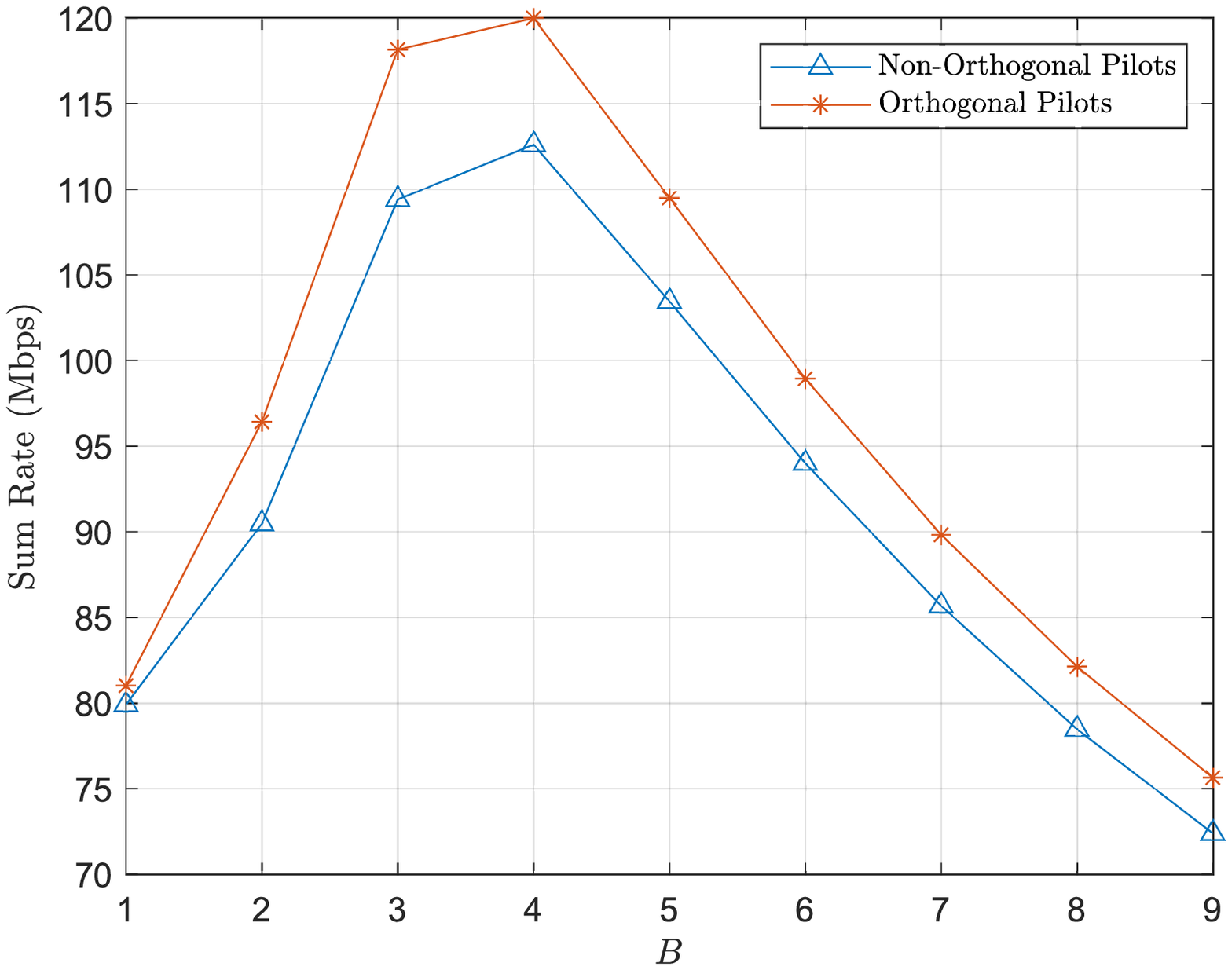}
\fi
\caption{Sum rate vs. number of transmission slots for scheduled random access in slow-fading scenario with $K=150$, $M=64$, $N=2000$, $T=2000$ and $L=600$. ``Non-orthogonal pilots'' refers to using AMP for joint activity detection and channel estimation. ``Orthogonal pilots'' refers to using the covariance approach for activity detection with $L_1=200$, followed by assigning orthogonal pilots of $L_2=400$ for channel estimation.}
\label{fig:rate_amp_vs_cov_sch_overload}
\end{figure}

Under the above parameters, we first show an example of optimizing $B$ assuming 
a value of $L=600$; subsequently, we present simulations that also optimize over $L$. 
Fig.~\ref{fig:rate_amp_vs_cov_sch_overload} shows the user sum rate versus the
number of scheduled slots $B$ assuming that the MMSE receive beamforming is
used at the BS for the case of $L=600$. 
In this overloaded system both the cases of using the orthogonal and
the non-orthogonal pilots for channel estimation benefit from having 
$B > 1$ scheduled slots for data transmission.
It can be observed that the optimal $B$ occurs when the system is close to
fully loaded and $\frac{K}{MB} < 1$.

\begin{figure}[t]
\centering
\ifdefined\ONECOLUMN
\includegraphics[width=0.6\columnwidth]{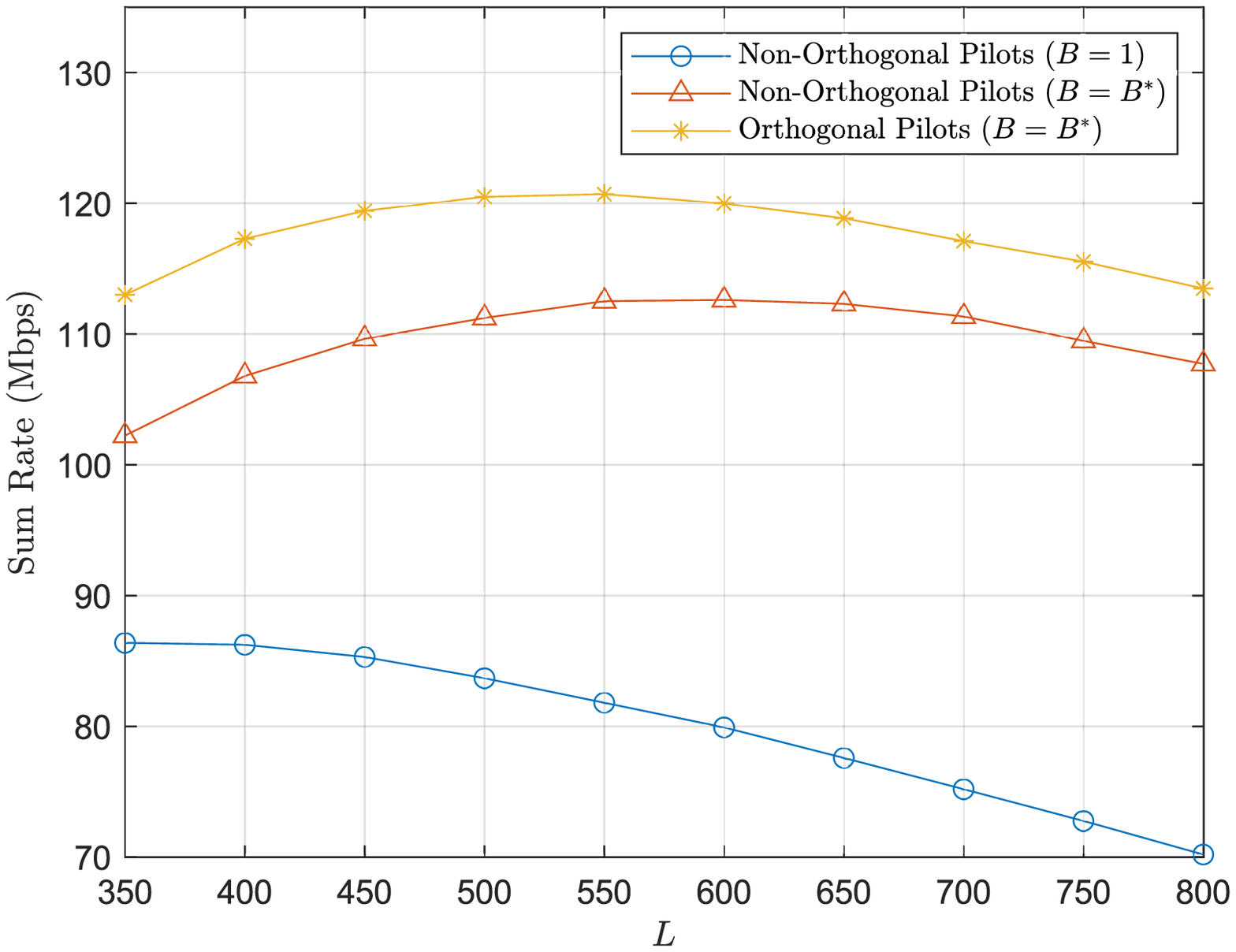}
\else
\includegraphics[width=\columnwidth]{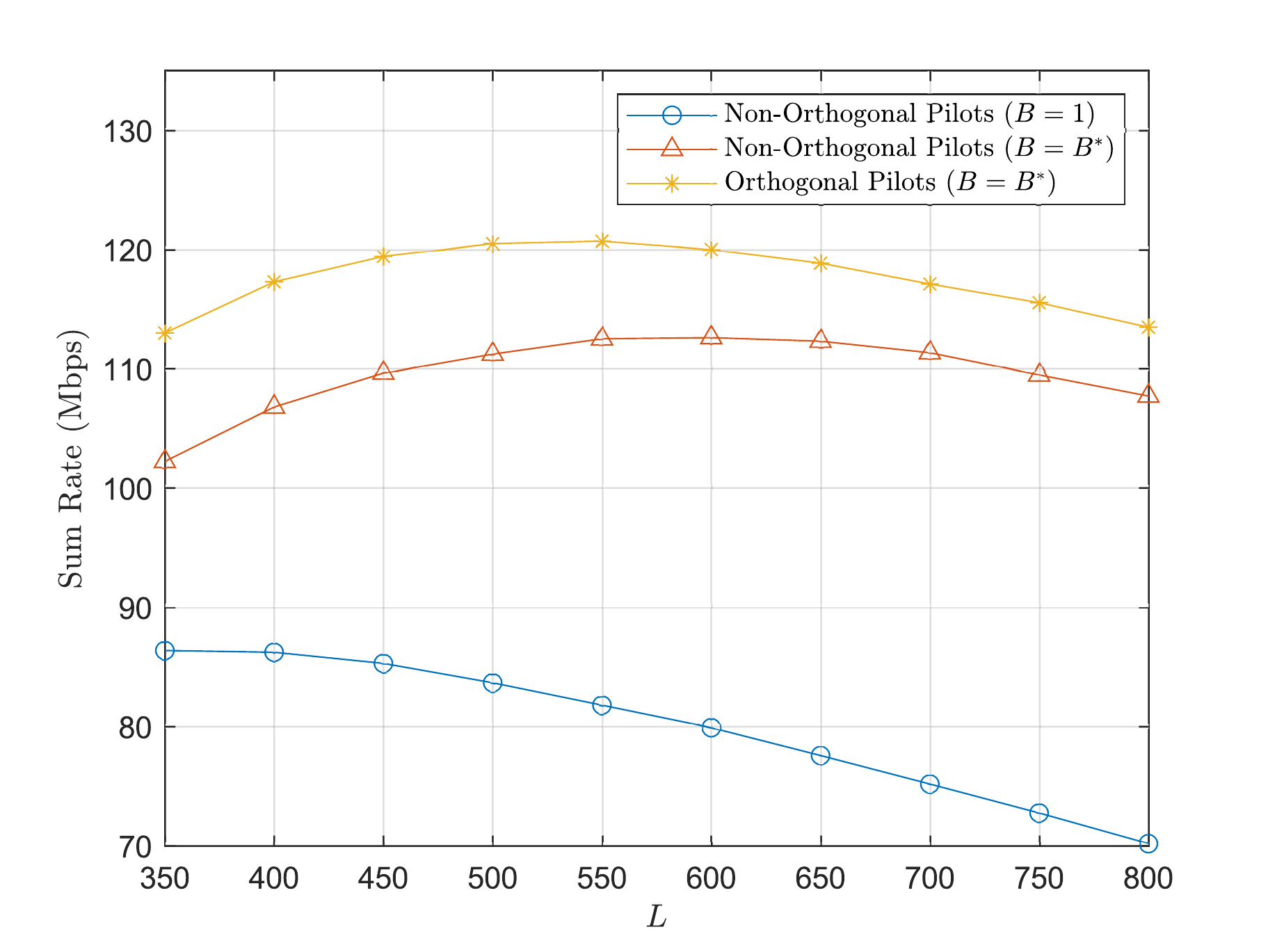}
\fi
\caption{Sum rate vs. pilot length for scheduled random access in slow-fading scenario with $K=150$, $M=64$, $N=2000$, $T=2000$. ``Non-orthogonal pilots'' refers to using AMP for joint activity detection and channel estimation. 
``Orthogonal pilots'' refers to using the covariance approach for activity detection 
with $L_1=200$, following by assigning orthogonal pilots of length $L-L_1$ for 
channel estimation. Where indicated, users are scheduled in the optimal number of $B^*$ 
transmission slots.}
\label{fig:rate3}
\end{figure}

Fig.~\ref{fig:rate3} compares the sum rate of the three approaches for
scheduled random access discussed in this section plotted against the total
pilot length $L$, where for each value of $L$, we numerically find the optimal
$B$ (when applicable) and use this $B^*$ when evaluating the sum rate.  
These numerical results are summarized along with the required feedback in
Table~\ref{table:sum_rate}.
Note that increasing the length of the pilots improves the channel estimation error and increases the achievable rate for the remaining symbols but decreases the number of slots available for data transmission, thus creating a trade-off. We observe that for joint activity and channel estimation using non-orthogonal pilots via AMP without feedback ($B=1$), there is little gain from using longer pilots. This is because the bottleneck is the significant interference due to having all users transmit simultaneously in an overloaded system. 
The sum rate can be significantly improved by 
scheduling users over an optimized number of $B^*$ slots. Furthermore, when orthogonal pilots are assigned to the users, we see a higher sum rate due to the improved channel estimation from the use of orthogonal pilots.  

These benefits come at only a small cost of feedback as summarized in 
Table~\ref{table:sum_rate}.
For the case of assigning orthogonal pilots and the scheduling slots to the $K$
users, the amount of feedback required is $K\log(e)$, or roughly $216$ bits per
frame, and this remains the same regardless of $B$. This works out to be $108$ kbps. 
If non-orthogonal pilots are used for channel estimation, by referring to the 
fundamental bounds in \cite{kang2020minimum}, the minimum feedback needed to 
assign users to $B^*=4$ slots is only $18$ bits per frame of feedback, or $9$ kbps.

\ifdefined\ONECOLUMN

\begin{table}
\centering
\caption{Sum Rate and Feedback Cost for Scheduled Random Access Schemes in Slow Fading}
\label{table:sum_rate}
\resizebox{!}{!}{%
\begin{tabular}{c c c}
\toprule
Algorithm & \makecell{Feedback (kbps)} & \makecell{Sum Rate (Mbps)} \\ [0.5ex] \midrule
\makecell[l]{Joint Activity and Channel Estimation with \\ Non-Orthogonal Pilots Using AMP ($B=1$)} & 0 & 87\\ [2ex] \midrule
\makecell[l]{Joint Activity and Channel Estimation with \\ Non-Orthogonal Pilots Using AMP ($B=B^*$)} & 9 & 112\\ \midrule
\makecell[l]{Covariance Method for Activity Detection and \\ Orthogonal Pilots for Channel Estimation ($B=B^*$)}& 108 & 121\\
\bottomrule
& & \\
\end{tabular}
}
\label{tab:slow_fading}
\end{table}

\else

\begin{table}
\centering
\caption{Sum Rate and Feedback Cost for Scheduled Random Access Schemes in Slow Fading}
\label{table:sum_rate}
\resizebox{!}{!}{%
\begin{tabular}{c c c}
\toprule
Algorithm & \makecell{Feedback\\(kbps)} & \makecell{Sum Rate\\(Mbps)} \\ [0.5ex] \midrule
\makecell[l]{Joint Activity and Channel \\Estimation with Non-Orthogonal \\Pilots Using AMP ($B=1$)} & 0 & 87\\ [2ex] \midrule
\makecell[l]{Joint Activity and Channel \\Estimation with Non-Orthogonal \\Pilots Using AMP ($B=B^*$)} & 9 & 112\\ \midrule
\makecell[l]{Covariance Method for Activity \\Detection and Orthogonal Pilots\\ for Channel Estimation ($B=B^*$)}& 108 & 121\\
\bottomrule
& & \\
\end{tabular}
}
\label{tab:slow_fading}
\end{table}

\fi

\section{Conclusion} \label{sec:conclusion}

This paper investigates the benefit of \emph{scheduling} for massive random
access in massive MIMO systems---an approach made possible by recent
advancements in efficient activity
detection and feedback. We propose a three-phase scheduled random access
procedure that begins with activity detection using non-orthogonal pilots,
followed by BS feedback for scheduling, and finally, data transmission for the
scheduled users.  User activities are detected from the non-orthogonal pilots
using either the AMP algorithm, which also provides a channel estimate, or via the covariance approach.
Scheduling is implemented via a single common feedback message from the BS to
the active users. 
Specifically, we quantify the cost of feedback needed for scheduling, and point
out that the feedback message can be used to schedule users to distinct transmission
slots to avoid collision, as well as to assign orthogonal pilots to different users
for channel estimation.  Leveraging the results in \cite{kang2020minimum}, we
show that the minimum feedback rate to ensure collision-free scheduling can be very low. 

We investigate the performance of the proposed scheduled approach to massive random access in comparison to contention-based or non-scheduled approaches in both the fast-fading setting where activity detection and data transmission occur in different coherence blocks, and the setting where fading is slow relative to the latency requirements of the system, so activity detection and data transmissions occur within a single coherence block.

In the fast-fading setting, the proposed scheduled random access approach is compared to the uncoordinated coded-ALOHA-based CPA approach. We show that the use of activity detection and feedback to enable scheduling can lead to notable improvements in system performance, such as the increased average number of transmissions per slot and the decreased overall power consumption, at a cost of only a small amount of feedback.
In the slow-fading setting, the proposed scheduled approach is compared with the approach of joint activity detection and channel estimation via AMP \cite{Liu2018}. 
We show that in an overloaded system, scheduling users to different transmission slots requires a very small amount of feedback. In addition, assigning orthogonal pilots to the users for channel estimation can lead to further improvement in sum rate at a moderate cost of feedback.

 These results establish that the use of relatively small amounts of feedback in random access protocols can lead to significant gains in system performance and efficiency, indicating that feedback should be considered in the development of future random access protocols.

\bibliographystyle{IEEEtran}
\bibliography{IEEEabrv,references}

\end{document}